\documentclass[12pt,letterpaper]{article}
\usepackage[utf8]{inputenc}
\usepackage{comment}
\usepackage{cite}
\usepackage[allow-number-unit-breaks]{siunitx}
\usepackage{amsmath,amssymb,amsfonts}
\usepackage{mathtools}
\usepackage{relsize}
\usepackage[framemethod=tikz]{mdframed}
\usepackage[toc,page]{appendix}

\usepackage{tabularx}
\usepackage{algorithmic}
\usepackage{graphicx}
\usepackage{textcomp}
\usepackage{xcolor}
\usepackage{xfrac}
\usepackage{lipsum}
\usepackage{adjustbox}
\usepackage{float}
\usepackage{placeins}
\usepackage{wrapfig}
\usepackage{ar}
\usepackage{caption}
\usepackage{subcaption}
\usepackage[utf8]{inputenc}
\usepackage{url}
\usepackage{authblk}

\usepackage{tikz}       
\usepackage{pgfplots}
\pgfplotsset{compat=1.18} 
\usetikzlibrary{positioning, bending, shapes, fit, arrows, arrows.meta}

\begin{document}

\title{{\raggedright
Precision Air Flow Control via EHD Actuator: A Co-simulation and Control Design Case Study
}}

\author[1]{Afshin Shaygani\thanks{Corresponding author: ashaygan@uwo.ca}}
\author[2]{Kazimierz Adamiak}
\author[3]{Mehrdad R. Kermani}

\affil[1,2,3]{Department of Electrical and Computer Engineering, Western University, London, Ontario, Canada N6A 5B9}
\affil[1]{ \texttt{ e-mail: ashaygan@uwo.ca}}

\renewcommand\Authands{ and }

\maketitle

\begin{abstract}

A Dielectric Barrier Discharge (DBD) plasma actuator for controlling airflow is proposed. It consists of diverging and converging nozzles, two concentric cylinders and an actuator mounted in-between the two cylinders. The actuator employs electrohydrodynamic (EHD) body force to induce an air jet within the air gap between the two cylinders, effectively creating a suction area while passing through the diverging nozzle, due to the Coanda effect. While merging with the air stream inside the inner cylinder, the Coanda jet effectively enhances amplification of the airflow. The outflow rate is measured by a velocity sensor at the outlet and controlled by the plasma actuator. The control strategy is based on the Active Disturbance Rejection Control (ADRC) and compared to the baseline PID controller. The actuator was modelled by seamlessly linking two modeling platforms for a co-simulation study. The CFD simulation of the plasma and airflow was carried out in the COMSOL multi-physics commercial software, and the control was implemented in the Simulink. The DBD plasma model was based on the two-species model of discharge, and the electric body force, calculated from the plasma simulation, was used in the Navier-Stokes equation for the turbulent flow simulation. The plasma-air flow system was analyzed using the input (the actuator voltage) and output (the outlet flow rate) data for the control design. Finally, the performance of the system of air flow control device was tested and discussed in the co-simulation process.

\end{abstract}

{\setlength{\parindent}{0cm}
}

\section{Introduction}

Non-equilibrium plasma produced by electric discharges, has demonstrated its effectiveness in diverse applications across a broad spectrum of fields, such as medicine, science, and engineering. Most common types of discharges used in such fields include corona and Dielectric Barrier discharges (DBD), which are self-sustained, low energy electric discharges. They exhibit different modalities at different polarities and voltage levels. The plasma is formed within a controlled environment by applying a significantly high potential difference between two conductive electrodes. When a dielectric material, known as a barrier, is positioned along the path of discharge between the electrodes, and the electrodes are supplied with AC voltage, the resulting discharge is termed as Dielectric Barrier Discharge (DBD). On the other hand, if there is no insulating barrier present and a DC voltage is supplied, the resulting discharge is known as a corona discharge. These types of discharges find applications in biomedicine, surface activation and disinfection, chemical compound decomposition \cite{stryczewska2022applications,li2019review,zeghioud2020review}, as well as in the electrohydrodynamic (EHD) flow actuators \cite{shaygani2023mean,hehner2019stokes,moralev2020localized,vernet2018flow}, to name a few.

EHD flow actuators demonstrate remarkable effectiveness in a range of engineering applications, such as EHD thrusters, pumps, heat transfer enhancement \cite{wang2023local,zeng2023negative}, boundary layer control devices \cite{hehner2020virtual,xu2019dielectric}, thermal management of electronics \cite{wen2022recent} and air amplification systems \cite{rubinetti2023electrohydrodynamic,rubinetti2023silico}. They relay on the heat, mass and momentum transfer processes. DBD plasma actuators are safer, more reliable and efficient compared to corona actuators, and therefore, used more often \cite{unfer2010modeling,moreau2007airflow}. DBD actuators are categorized into three distinct classes based on the applied voltage waveform. An alternating voltage is used in the first category, in which the actuation is in the steady mode of operation. In the second category, the alternating voltage provided undergoes modulation, leading to actuation in an unsteady or burst mode of operation. In the third category, actuation is achieved using a nanosecond pulse waveform \cite{unfer2010modeling}. In the first and second categories, the conversion of electrical energy into kinetic energy is driven by the momentum transfer from charged species to neutral molecules, and, therefore, for the induced EHD flow. The working principle in the third category is based on the generation of very steep thermal, pressure, viscosity, and density gradients and on the energy transfer from moving shock waves to the background flow \cite{correale2015energy}. This type of actuation is used mostly at high Reynolds and Mach numbers. Most recent reviews on DBD plasma actuators and their applications can be found in \cite{wen2022recent,peng2023review,johnson2017recent}.

Air amplification systems using corona discharge actuators has been recently studied by Rubinetti et al. \cite{rubinetti2023electrohydrodynamic,rubinetti2023silico}, where the flow rate can be controlled by increasing the supply voltage to the corona discharge actuator. Their proposed design leverages the Coanda effect \cite{panitz1972flow,dumitrache2012mathematical} to enhance the air flow rate by enforcing the air flow to remain attached to the Coanda jet. The Coanda effect is a phenomenon where a fluid stream tends to stay adhered to an adjacent solid surface. It is used as the underlying principle for the flow rate amplification technique. The maximum voltage level is constrained by the occurrence of electrical breakdown in air. However, by employing a DBD plasma actuator, a promising alternative with fewer limitations, the potential for achieving higher flow rates becomes feasible. This applies to both fluid mechanics and high-voltage engineering perspectives, enabling the development of a safer actuator design. Surprisingly, despite these advantages, comprehensive studies on its performance in air amplification have not been conducted, to the best of our knowledge.

The evaluation of device performance, such as EHD actuated air flow rate enhancement systems, EHD micro pumps and other devices utilizing EHD actuators, can be approached from multiple viewpoints, including fluid mechanics, electrostatics, and control system design perspectives. From the electrostatics point of view, several factors such as electrical breakdown, EHD body force generation and power consumption are considered as design criteria. The fluid mechanics point of view evaluates the device in terms of the flow rate, flow velocity profile, pressure variations, and the fluid flow regimes in which the actuator operates. Within engineering applications, EHD actuators offer a flexible solution for the design and implementation of a wide range of control strategies, owing to their exceptional actuation speed.

The control system evaluation approach is based on the response of the fluid flow to the actuation mechanism. While it was shown that the flow response close to the plasma actuator can be locally approximated with low order linear-time-invariant systems \cite{pereira2015analysis}, non-local behavior of the system can be quite complicated and is affected by the actuator specifications, such as the geometry, hysteresis effects \cite{benard2011benefits}, and the flow regimes. The choice of the control strategy varies depending on the device and specific applications at hand. For DBD plasma-based actuation, for example for flow separation control in weak turbulence or transient regimes, reinforcement learning methods have been considered \cite{shimomura2020closed, shimomura2020experimental,wang2022deep,vinuesa2022flow}. Conventional methods of closed-loop feedback control based on the PID controllers have also been extensively used
\cite{kwan2019pedagogical,kriegseis2013closed}. While there is a significant body of literature focusing on EHD-based actuation for flow control around airfoils, there is a notable lack of research on its application in internal flows, such as EHD micro-pumps and air enhancement systems.

Active disturbance rejection control (ADRC) and PID are among the most widely used control methods for industry applications. However, the ADRC method introduces a novel approach that actively estimates and rejects disturbances, setting it apart from traditional control methods. It has demonstrated successful applications in various fields, including motion control, flight control, power converters, robotics, fuel cell systems, chemical processes, superconducting RF cavities, axial flow compressors, and micro-electro-mechanical systems. First proposed by Han \cite{han2009pid}, ADRC has proven effective in cases where accurate models are lacking or when dealing with model uncertainty. The latest assessments of the ADRC technique can be found in publications \cite{zheng2018active,feng2017active}, while foundational research on the subject is presented in \cite{herbst2013simulative,chen2011tuning,herbst2021transfer}.

In the field of engineering, the simulation of real-world complex systems, especially those that are hybrid and governed by multiple physics, presents unique challenges. These systems often constitute a small part of larger interconnected systems, where they are controlled by and interact with other components of the overall system. To effectively model such intricate systems and implement integrated control strategies seamlessly, co-simulation emerges as a valuable approach. Co-simulation enables the exchange of data between multiple software systems, called Functional Mock-Up Unit (FMU) \cite{blochwitz2011functional,blockwitz2012functional,hatledal2019language}, allowing each system to solve problems independently while adhering to their respective solver definitions. This integration leverages software simulation programming languages to connect loosely coupled stand-alone subsystems, enabling a comprehensive simulation of the entire system. By facilitating effective communication between subsystems at predefined time intervals, co-simulation offers a holistic approach to analyze complex engineering systems, providing insights into the intricacies of future designs and ensuring seamless integration of control strategies within the larger system \cite{gomes2017co}. The commercial software package Comsol LiveLink for Simulink \cite{COMSOL} enables such co-simulation modeling.

The presented study utilizes a co-simulation approach to model the airflow dynamics in a DBD actuated control system. Drawing from the previously validated two-species model \cite{shaygani2021dielectric}, the research takes into account continuity equations for both charge species, the Poisson equation for the electric field, and the surface charge continuity. For fluid dynamics, the NS equations are employed, focusing on conservation of mass and momentum for neutral molecule movement, assuming an incompressible, turbulent flow using \(k-\omega\) model, and without thermal considerations. The research aims to control the airflow rate at the device outlet by adjusting the voltage supplied to the actuator. Additionally, the proficiency of two distinct control strategies, PID and ADRC, is evaluated, with an emphasis on path tracking and disturbance rejection capabilities.

Addressing the need for precise airflow control, this research emphasizes modulating the airflow rate at the outlet using the AC-DBD plasma actuator. A co-simulation technique provides insights into the interactions between plasma and fluid flow, laying the groundwork for effective control strategy designs. A comparison is made between the Active Disturbance Rejection Control (ADRC) and the conventional PID controller to pinpoint optimal flow rate regulation at the outlet. The study broadens its scope through examining the impact of various parameters on flow rate and transitions into closed-loop control for consistent airflow regulation. While the study predominantly relies on numerical methods, it paves the way for subsequent experimental validations and builds a foundation that future research can expand upon. Consequently, this work offers insights and guidance for both researchers and engineers, advancing the understanding in the field of airflow dynamics control.

\section{Problem formulation}

\subsection{Model description}

\begin{figure}[htbp]
\centerline{\includegraphics[width=0.80\textwidth]{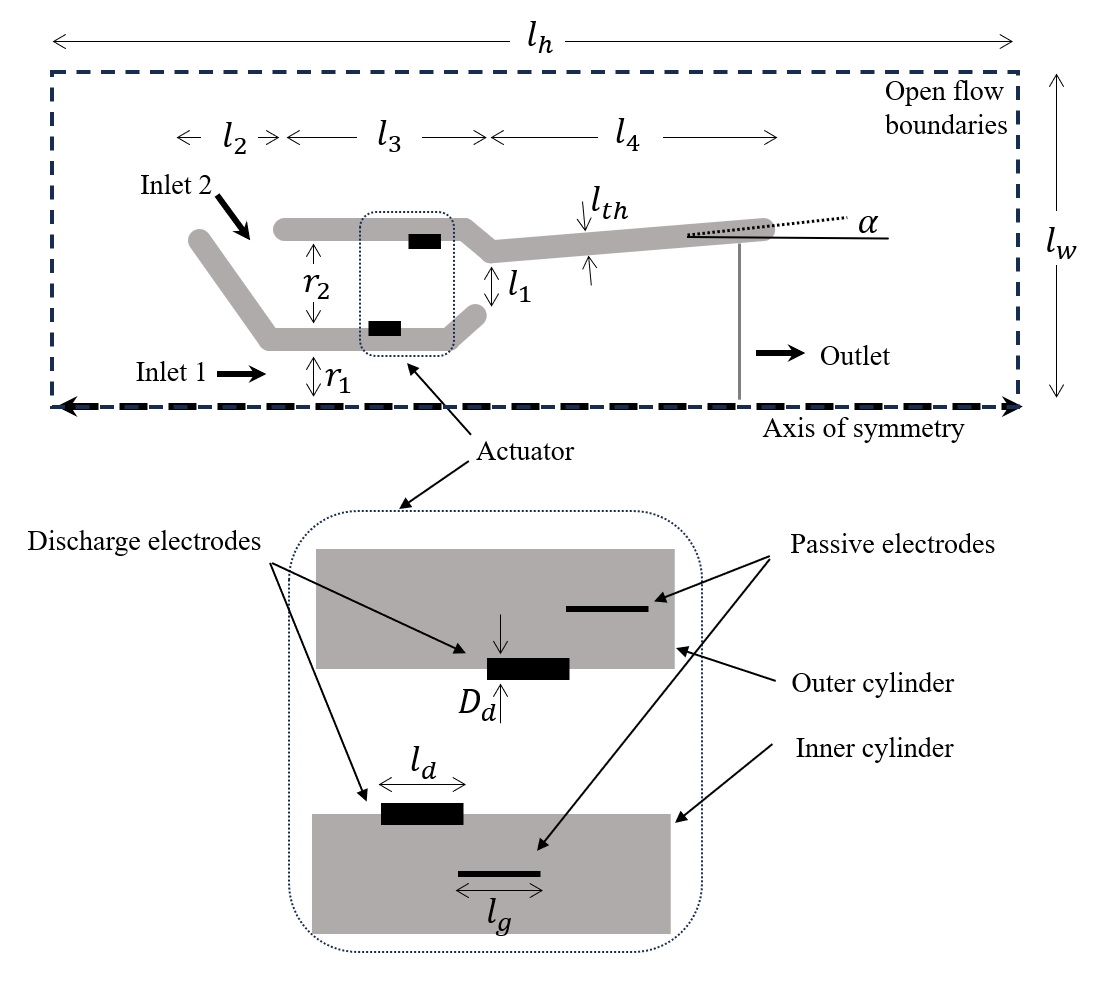}}
\caption{The airflow control device(gray) and the DBD plasma actuator (black) shown inside the computational domain. The empty space indicates the airflow region.}
\label{schematic}
\end{figure}

The airflow control system equipped with a DBD actuator, studied in this paper, is shown in Figure \ref{schematic}. The system comprises two concentric cylinders, with radii \(r_1\) and \(r_1 + r_2\). There are two inlets for air to enter, and an outlet to exit. The actuator pulls in air through the suction area, creating a jet as it passes through the narrow gap of thickness \(l_1\) between the two cylinders. This jet in turn, induces a secondary flow from the main inlet, resulting in an amplification of airflow. Alternatively, the jet and the induced secondary flow can separate into two streams, one leaving the outlet, and another returning back to the central inlet and causing attenuation of airflow. The lengths of the embodiment holding the actuator, including the converging inlet, the co-centric cylinders, and the diverging outlet are \(l_2\), \(l_3\), and \(l_4\), respectively, with the diverging angle \(\alpha\).

The actuator consists of two separate DBD systems of ring-to-ring electrode type device. One is mounted on the inner cylinder and the other on the outer cylinder. Two discharge electrodes of thickness \(D_d\) and length \(l_d\) are exposed, and supplied with high sinusoidal alternating voltages with an amplitude \(V_{ap}\) and frequency \(f\). Two passive electrodes of length \(l_2\) (the ground electrodes) are encapsulated inside the body of the apparatus made of a dielectric material with the relative permittivity \(\epsilon_r\). The surface conductivity of the dielectric is assumed zero, and its thickness is \(l_{th}\). The space occupied by dry air with a relative permittivity of \(\epsilon_{air}\). The discharge problem is solved first, and then the period-average electrohydrodynamic body force \(F_{ehd}\) is calculated for the flow problem. Both problems are considered 2D-axisymmetric featuring the  computational dimension of \( l_h \times l_w\). Air is kept at the temperature of \(\SI{300}{K}\) and atmospheric pressure. The parameters of the model under investigation are detailed in Table \ref{tab1}.

\subsection{Mathematical model}

The mathematical model includes the electric discharge, the air flow and the control system. The discharge is modeled considering the two-species model with generic charge species: positive and negative \cite{shaygani2021dielectric}. Two charge transport equations describe the drift, diffusion, and annihilation of charge species under the influence of the electric field, surface and space charges. The electrostatics equations include Poisson equation, and conservation of charge due to the charge accumulation on the dielectric surface. The EHD body force, given by the product of the net space charge and electric field, is introduced to the incompressible NS equations for the fluid flow simulation. The space charge is assumed to be unaffected by the fluid flow. Therefore, a one-way coupling is considered between the charge transport and NS equations which effectively models the influence of EHD body force on the fluid flow. The flowrate at the outlet is measured, and used by the control law,  governed by a set of five ordinary differential equations (ODEs), to predict the actuator input voltage.

\subsubsection{Discharge model}

\paragraph{Governing equations}  \label{GE1}

The electric potential \(V\) is calculated by solving the Poisson equation:

\begin{equation}\label{ePo}
\nabla^2V=-\frac{\rho}{\epsilon} \\
\end{equation}
where \(\rho\) represents the net space charge density and \(\epsilon\) signifies the permittivity of air. The intensity of the electric field, \(E\), is derived as the gradient of the electric potential, expressed as \( \vec{E} = - \nabla V \). If \(c\) is the number density of charge species, the net space charge \(\rho\) is calculated as:

\begin{equation}\label{eSp}
\rho=\rho_p-\rho_n =  e\cdot (c_p - c_n ) 
\end{equation}
where \(e\) denotes the electronic charge, with the subscripts \(p\) and \(n\) indicating positive and negative ions, respectively. Charge species present or produced within the computational domain either participate in ionic reactions and become neutralized or are destroyed, or they exit the domain.

A series of charge continuity equations is solved to determine the number density \(c\) of charge species
\cite{shaygani2021dielectric}:

\begin{equation}\label{e3}
\begin{cases}
\frac{\partial c_p}{\partial t}+\nabla\cdot\left( c_p\mu_p\vec{E}-D_p\nabla c_p\right)=-\beta_{np} c_p c_n\\

\frac{\partial c_n}{\partial t}+\nabla\cdot\left(- c_n\mu_n\vec{E}-D_n\nabla c_n\right)=-\beta_{np} c_p c_n
\end{cases}
\end{equation}

where \(t\) is the time, \(\mu_p\) and \(\mu_n\) are the ion mobilities, equal to \(2.43\times{10}^{-4}\;\sfrac{m^2}{\left(V\cdot s\right)}\) and \(2.7\times{10}^{-4}\;\sfrac{m^2}{\left(V\cdot s\right)}\) for the positive and negative ions, respectively \cite{georghiou2005numerical}, \(D\) is the diffusivity of charge species which is equal to \(4.6\times{10}^{-6}\;\sfrac{m^2}{s}\), and \(\beta\) is the recombination coefficient which is equal to \(2.0\times{10}^{-13}\;\sfrac{m^3}{s}\)  \cite{georghiou2005numerical}.

Charge deposition on the surface of the dielectric is described by conservation of charge law. Neglecting the conductivity effect and surface reactions, this equation reads as \cite{shaygani2021dielectric}:

\begin{multline}\label{e4}
\frac{\partial\rho_s}{\partial t} = {\vec{n}}_\parallel \cdot {\vec{J}}_\parallel + {\vec{n}}_\perp \cdot {\vec{J}}_\perp =  e\cdot E_\perp\cdot 
  \begin{cases}
    \mu_p\cdot c_p       & \quad \text{if } E_\perp < 0 \\
    \mu_n\cdot \ c_n  & \quad \text{if } E_\perp >0
  \end{cases}
 \end{multline}
where \(\rho_s\) denotes the surface charge density, \({\vec{n}}_\perp\) and \({\vec{n}}_\parallel\) represent the normal and tangential vectors to the dielectric surface, respectively. \({\vec{J}}_\perp\) and \({\vec{J}}_\parallel\)  are the normal and tangential components of the drift current density above the surface and the conduction current density along the surface, respectively. \(E_\perp\) is the electric field normal to the dielectric surface. The surface conductivity has been neglected, therefore the conduction current is \(0\) in this case.

The continuity equation expresses the discontinuity of the electrostatic displacement vector resulting from the surface charge at the interface between air and dielectric. It is given by: 

\begin{equation}\label{e5}
\vec{n}\cdot\left(\varepsilon\vec{E_d}-\varepsilon\vec{E_a}\right)=\rho_s
\end{equation}
where \(\vec{E_d}\) and \(\vec{E_a}\) denote the electric fields within the dielectric and air, respectively. Those charge species who don't take part in the deposition process described by the charge conservation and continuity equations, either leave the domain through the surrounding boundaries, or recombine with the opposite charges.

Due to computational constraints, the ionization process is disregarded in the two-species discharge model presented in this study. As a result, the boundary conditions governing the charge concentrations at the active electrode, as formulated by Kaptzov's hypothesis, are adopted based on Peek's formula: 

\begin{equation}\label{e6}
E_s=3.1\times{10}^6\cdot\ p\cdot\left(1+\frac{0.308}{\sqrt{r\cdot p}}\right)\left[\frac{V}{m}\right]
\end{equation}
where \(E_s\) denotes the electric field on the surface  of the discharge electrode, \(r\) is its radius in cm, and \(p\) is the gas pressure. For active electrode voltages exceeding the onset level, Kaptzov's hypothesis states that \(E_s\) remains constant.

\paragraph{Initial and boundary conditions} 

The simulation begins with the assumption of charge neutrality, therefore number charge densities are initialized to zero. The appropriate boundary conditions, as listed in Table \ref{bc}, are employed for the electrostatics equations. On the discharge electrode,\(V_{ap}\) is assumed, while the passive electrode is grounded. For the dielectric surface, Eq. (\ref{e5}) is utilized, and the far-field air boundaries are assigned a zero charge condition. Regarding species transport, a charge injection law is enforced at the tip of the discharge electrode, while the rest of the boundaries follow the outflow condition, where the convective flux is set to zero. In the ionization region, charge species are produced, a process effectively represented by three-species models \cite{shaygani2023mean,shaygani2023self}. The two-species model employed in this study, however, lacks the capability to simulate this specific region. Nevertheless, given the ionization region's minimal size, particularly near the discharge electrode's surface, it can be disregarded for computational efficiency. Consequently, to accurately depict the discharge using a two-species model, we implement an injection law that compensates for the charge species production that the model intrinsically does not capture. The charge injection law determines the charge values (\(c_+\) and \(c_-\)) at the tip based on equations (\ref{law1}) and (\ref{law2}). Here, \(\beta_1\) and \(\beta_2\) represent fixed experimental parameters chosen to ensure that \(E_\perp\) closely approximates \(E_s\).

\begin{equation}\label{law1}
c_+=
\begin{cases}
    max(0,\beta_1(\|E_\perp\|-E_s) + \beta_2 c_p)  & \quad  E_\perp > 0 \\
    0  & \quad  E_\perp <0
  \end{cases}
\end{equation}

  \begin{equation}\label{law2}
c_-=
\begin{cases}
    0  & \quad  E_\perp > 0 \\
    max(0,\beta_1(\|E_\perp\|-E_s) + \beta_2 c_n)  & \quad  E_\perp <0
  \end{cases}
\end{equation}

\begin{table}[htbp]
\begin{center}

\caption{Discharge boundary conditions}
\begin{tabular}{l l l l}
\hline
\hline
\\ [-5pt] 
\textbf{Equation}& \textbf{Discharge tip} & \textbf{Dielectric surface} & \textbf{Remote boundaries}
\\ [+2pt]
\hline
\\ [-5pt]
Poisson \((V)\)  & \(V_{ap}\) & (\ref{e5}) & \(\vec{n}\cdot(\varepsilon\vec{E})=0\) \\

Transport \((+)\) & \(c_+\) & \(\vec{n}\cdot(D_p\nabla c_p)=0\) & \(\vec{n}\cdot(D_p\nabla c_p)=0\)\\

Transport \((-)\) & \(c_-\) & \(\vec{n}\cdot(D_n\nabla c_n)=0\) & \(\vec{n}\cdot(D_n\nabla c_n)=0\)\\

\\ [-5pt]
\hline
\hline
\end{tabular}
\label{bc}
\end{center}
\end{table}

\subsubsection{Fluid flow model}

\paragraph{Governing equations}  \label{GE2} 

Under the assumption of dominant Coulomb force and a turbulent flow, the electric force  \(\vec{F_{ehd}}=\rho\cdot\vec{E}\) can be computed. This force can then be incorporated as a body force into the Navier-Stokes equations (NS), allowing for the simulation of electrohydrodynamic (EHD) flow. The incompressible NS equations govern the conservation of mass and momentum for neutral gas molecules:

\begin{equation}\label{eNS}
\begin{cases}
\frac{\partial \rho_a}{\partial t} + \nabla \cdot \left( \rho_a  \vec{u} \right) = 0  \  \xRightarrow{\text{incompressible}}  \   \nabla \cdot \vec{u} = 0 \\

\frac{\partial \vec{u}}{\partial t} + \vec{u} \cdot \nabla \vec{u} = -\frac{1}{\rho_a} \nabla p + \nabla \cdot \left[ (\nu + \nu_t) \nabla \vec{u} \right] + \vec{g} +  \frac{\Vec{F_{ehd}}}{ \rho_a }\\

\frac{\partial k}{\partial t} + \vec{u} \cdot \nabla k = \nabla \cdot \left[ (\nu + \sigma_k \nu_t) \nabla k \right] +\frac{P_k}{\rho_a} - \beta^* \omega k \\

\frac{\partial \omega}{\partial t} + \vec{u} \cdot \nabla \omega = \nabla \cdot \left[ (\nu + \sigma_{\omega} \nu_t) \nabla \omega \right] + \alpha \frac{\omega}{k \rho_a} P_k - \beta \omega^2 \\

P_k = \nu_t \left( \frac{\nabla \vec{u} + \nabla \vec{u}^T}{2} \right) : \left( \frac{\nabla \vec{u} + \nabla \vec{u}^T}{2} \right) 
\end{cases}
\end{equation}

where \( \rho_a\), \(p\), and \( \nu\) are the density, pressure, and kinematic viscosity of the air, respectively, \( \nu_t\) is turbulent viscosity, \(\vec{u}\) is the velocity vector of neutral air molecules, \(\Vec{g}\) is the vector of gravity, and \(\Vec{F_{ehd}}\) is the EHD body force. \(k\) and \(\omega\) represent the turbulence kinetic energy per unit mass, and the specific rate of dissipation of the turbulence kinetic energy, respectively. The turbulent model constants are \(\beta = 0.072\),  \(\beta^*=0.09\), \(\sigma_{\omega}=0.5\), \(\sigma_{k}=0.5\), and \(\alpha=0.52\), and \(P_k\) is the production of turbulence kinetic energy defined. The operator \( : \) denotes the double dot product. The kinetic energy due to the thermal energy is ignored, therefore, gas heating is not considered in this study. However, in certain applications, it is necessary to take into account the conversion of thermal energy into kinetic energy as well \cite{shaygani2023numerical}.

\paragraph{Initial and boundary conditions}  \label{Boundary conditions-F}

The simulation starts assuming the air is stationary, thus initializing the velocity vector to zero. On the solid surfaces, no-slip Dirichlet boundary condition (\(\Vec{u}=0\)) is considered. This includes the surface of the discharge electrode, and the body of the apparatus. On the remote boundaries, it is assumed that the airflow can freely enter or leave the domain. Hence, zero normal stress, or constant pressure, is applied to the boundaries. Small nonzero initial values are specified for \(k\) and \(\omega\) throughout the flow domain. At solid walls, \(k\) is set to zero, and \(\omega\) set to a very high value using a wall function, reflecting high rate of dissipation. Zero normal gradient conditions are set for both \(k\) and \(\omega\) on the axis of symmetry. More details on the initial and boundary conditions can be found in  \cite{COMSOL2023CFD}

\subsubsection{Control strategy}

The discharge model simulates the DBD actuator to compute the EHD body force that influences the fluid as a driving force. By incorporating the EHD body force into the NS equations, the induced EHD flow within the apparatus is determined. Subsequently, the flow rate \(Q_o\) of the exiting air is evaluated at the outlet and recorded as the system's output. The objective is enforcing \(Q_o\) to follow the prescribed flow rate \(Q_{sp}\). Hence, the airflow control system is conceptualized as a single-input single-output (SISO) system, with voltage \(V_{ap}\) as the input and flow rate \(Q_o\) as the output, operating under the control of a control system modelled in Simulink. 

In the context of this study, various challenges arise due to both external disturbances, such as surrounding air motions, and internal uncertainties, including flow fluctuations and transition regimes resulting from air actuation. To address these challenges, the main control strategy employed is active disturbance rejection control (ADRC), which has proven its effectiveness in managing both internal and external disturbances and uncertainties.

\begin{figure}[htbp]
\centerline{\includegraphics[width=1.0\textwidth]{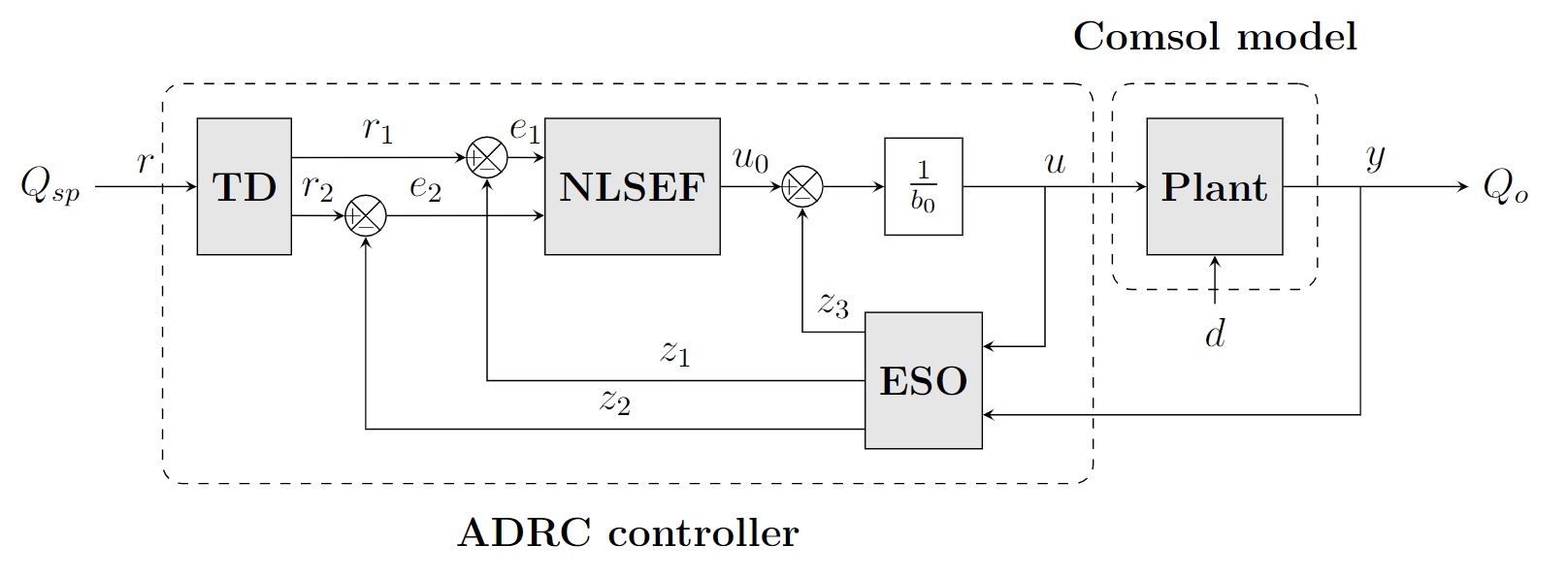}}
\caption{Structure of the \(2^{nd}\) order ADRC.} \label{chartC}
\end{figure}

In the ADRC methodology, the system dynamics including internal uncertainties and external disturbances, are transformed into a double-integral form using an extended state observer (ESO). Subsequently, the controller is constructed and designed based on this transformed representation, and two other main components, a tracking differentiator (TD), and a nonlinear state error feedback (NLSEF). The diagram in Fig. \ref{chartC} depicts the configuration of the second-order ADRC employed in this study.

\paragraph{Extended state observer} The ESO used in the structure of the ADRC is defined by the following set of ODEs:

\begin{equation}\label{ESO}
\begin{cases}

\epsilon_1 =  y -  z_1  \\
\epsilon_2 = -\text{fun}_1(-\epsilon_1,\alpha_1,\delta)  \\
\epsilon_3 = -\text{fun}_1(-\epsilon_1,\alpha_2,\delta) \\
\dot{z_1} = z_2 + \beta_{o1} \epsilon_1      \\
\dot{z_2} = z_3 + \beta_{o2} \epsilon_2 + b_0 u     \\
\dot{z_3} =   \beta_{o3} \epsilon_3      \\

\end{cases}
\end{equation}
where inputs to ESO are the control signal \(u=V_{ap}\) and the output \(y=Q_o\), and the outputs of ESO are \(z_1\), \(z_2\), and \(z_3\). The observer parameters are \(\beta_{o1}\), \(\beta_{o2}\), \(\beta_{o3}\),  \(\alpha_1\), \(\alpha_2\), and \(\delta\). The nonlinear function \(\text{fun}_1\) is defined as:

\begin{equation}\label{fun1}
\text{fun}_1(x,\alpha,\delta) =
\begin{cases}

\frac{x}{\delta^{\left(1-\alpha\right)}} & \lvert x\rvert  \le \delta  \\

 \lvert x\rvert^\alpha \text{sign\((x)\)} &  \lvert x\rvert \ge \delta  \\

\end{cases}
\end{equation}

\paragraph{Tracking differentiator} The fastest tracking of the input reference signal \(r\) and its derivative is achieved by the following formula known as the tracking differentiator of \(r\) \cite{han2009pid}:

\begin{equation}\label{td}
\begin{cases}

\dot{r_1} = r_2  \\
\dot{r_2} =  -R \; \text{sign}( r_1  -  r  +   \frac{r_2 \lvert r_2 \rvert}{{2 R}} ) \\
\end{cases}
\end{equation}
where \(R\) is a tuning parameter of acceleration, set as a desired frequency filter. The \(sign\) function in this study has been replaced with the \(tanh\) function to enhance numerical stability.

\paragraph{Nonlinear state error feedback} The NLSEF integrates the tracking signal, its derivative from the TD, and the signal from ESO, to formulate the following control law:

\begin{equation}\label{CL}
 u_0=   \beta_1 \text{fun}_1(e_1,\alpha_1,\delta) + \beta_2 \text{fun}_1(e_2,\alpha_2,\delta)
\end{equation}
where \(e_1 = r_1-z_1\), \(e_2 = r_2-z_2\), and \(\beta_1\) and \(\beta_2\) are adjustable parameters. Therefore, the control command is obtained by \( u = (u_0-z_3)/b_0\).

\section{Simulation results}

\subsection{Numerical model}

The COMSOL commercial software \cite{COMSOL}, based on finite element method, has been used for the numerical simulation of the discharge and fluid flow. The discharge was modeled using two modules: the Electrostatics and the Transport of Diluted Species modules to solve the Poisson, the charge transport, and the surface charge accumulation equations. The time-averaged EHD body force \(F_{ehd}\) was calculated by the Domain Ordinary Differential Equations and Differential Algebraic Equations module (Domain ODEs and DEAs). All discharge equations are solved in a fully coupled manner. For solving the transport equations, a first-order scheme was use, and to stabilize equations primarily governed by first-order derivatives of concentrations, the streamline diffusion was adopted in the computation process. The electric field equations were solved using the second-order accurate interpolation scheme. For air flow simulations, the NS equations were solved with piece-wise linear discretization scheme for velocity and pressure (i.e \(P_{1} + P_{1}\) elements), using the Turbulent Flow in Fluids module. For both discharge and fluid flow simulations, adaptive time stepping method was adopted based on the implicit backward differentiation formulation (BDF).

\begin{figure*}[htbp]
   \centering
    
     \begin{subfigure}[b]{0.48\textwidth}
    \includegraphics[width=1\textwidth]{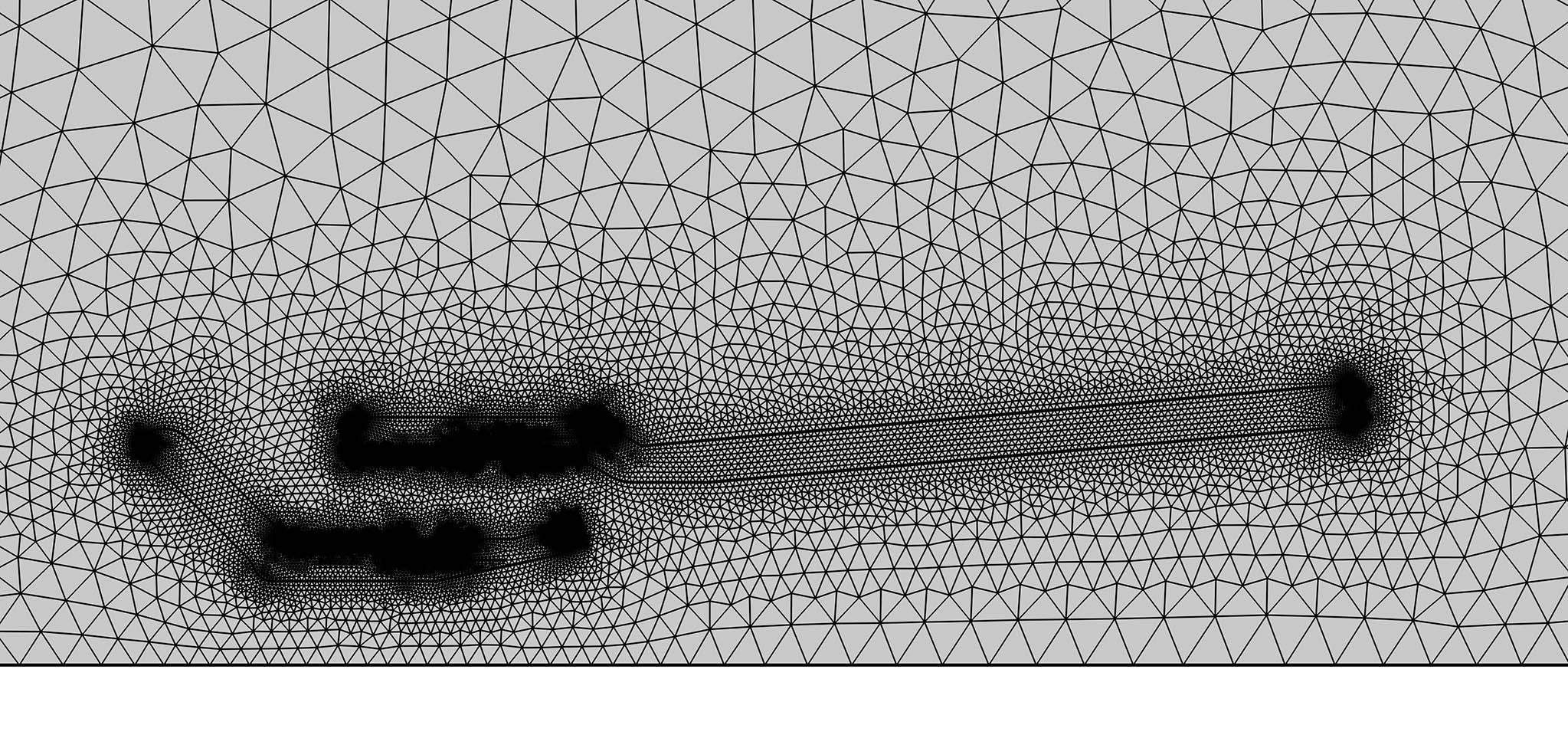}
    \caption{}
    \label{g1}
    \end{subfigure}
     \begin{subfigure}[b]{0.48\textwidth}
    \includegraphics[width=1\textwidth]{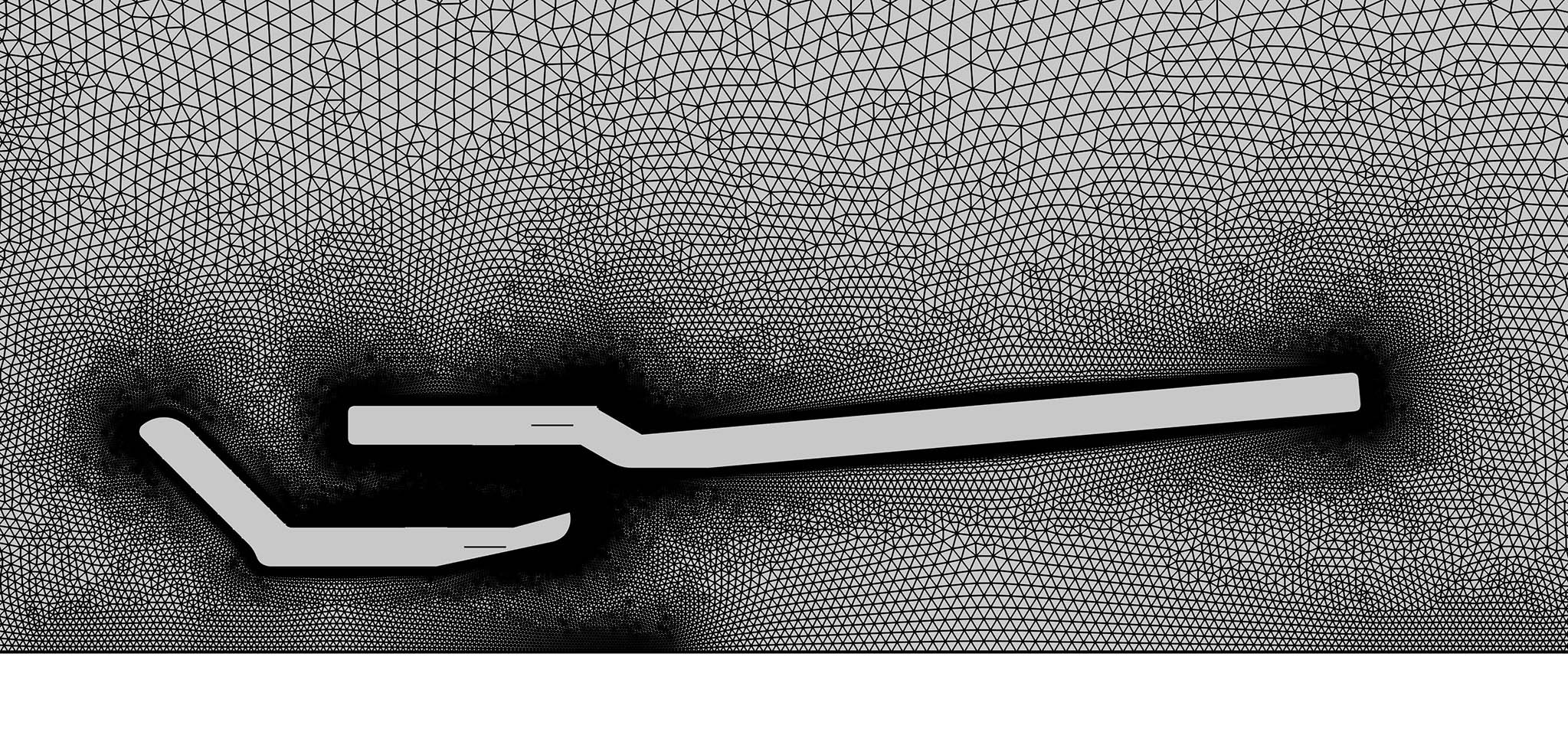}
    \caption{}
    \label{g2}
    \end{subfigure}
       
     \begin{subfigure}[b]{0.48\textwidth}
    \includegraphics[width=1\textwidth]{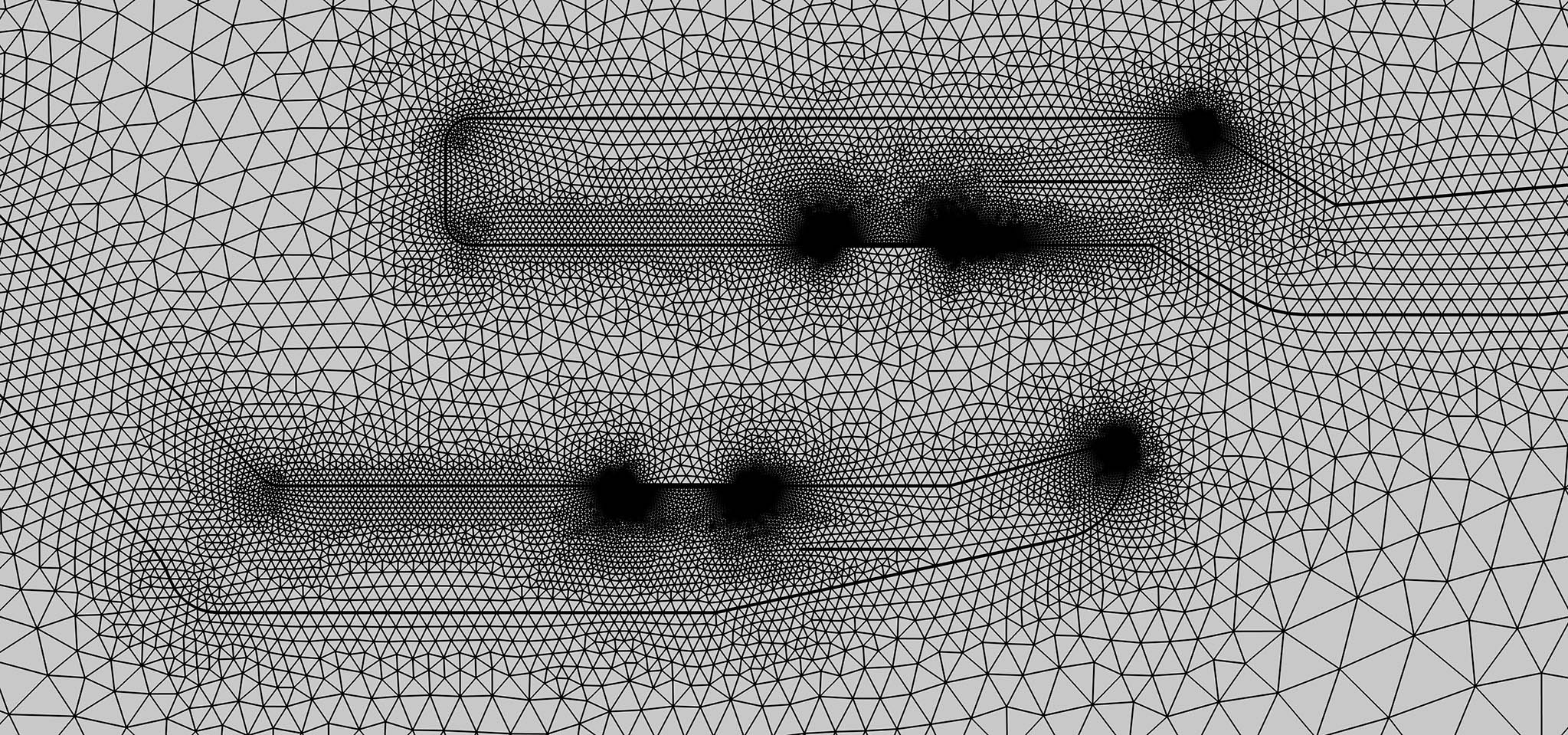}
    \caption{}
    \label{g3}
    \end{subfigure}
     \begin{subfigure}[b]{0.48\textwidth}
    \includegraphics[width=1\textwidth]{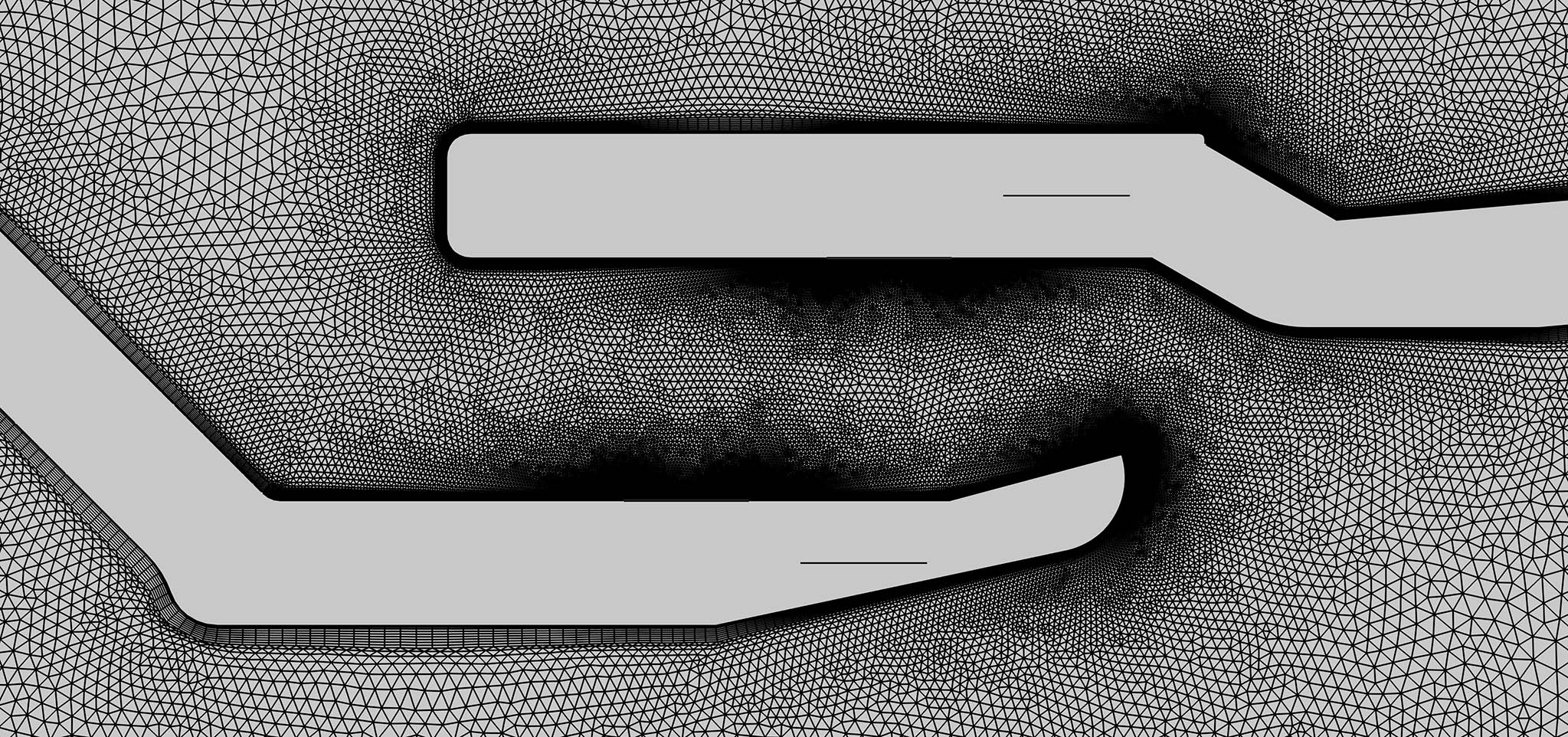}
    \caption{}
    \label{g4}
    \end{subfigure}
    
\caption{The unstructured triangular mesh and hybrid mesh employed for the discharge and air flow computations on the left and right columns, respectively. Near the actuator area, the grids were refined, shown in the second row.}
    \label{Grid}
\end{figure*}

An unstructured triangular mesh, and a hybrid mesh were used to discretize the computational domains for the discharge and fluid flow simulations with \(70,000\) and \(90,000\) elements, shown in figure \ref{Grid}, leading to \(1.1\) and \(2.3\) million degrees of freedom, respectively. The hybrid mesh comprised of unstructured triangular elements, far from the walls, and structured rectangular elements, near the walls. The refinement of elements was carried out in the discharge region. For the fluid flow simulation inside the cylinders and near the wall boundaries, element refinement was also performed. To ensure sufficient accuracy and reliable results, the solutions were assessed for grid independence using an iterative mesh refinement approach. The plasma simulation requires approximately 3 CPU hours, while the co-simulation for flow control takes about 35 CPU hours. These estimates are based on a system powered by an Intel Core i7-6700 CPU operating at 3.4 GHz.

\subsection{Model parameters and co-simulation settings}

\paragraph{Geometric parameters} 
The parameters of the air amplifier system studied in this paper are summarized in Table \ref{tab1}. The specifications not included in the table are: the angle of the converging nozzle -- \SI{45}{^{\circ}}, difference in axial positioning of the discharge electrodes -- \SI{8}{mm}, the axial distance between each pair of passive and active electrodes -- \SI{2}{mm}, thickness of the passive electrodes is negligible and irrelevant, and their radial positions are located in the middle of the dielectric material. A range of angles are considered for the diverging nozzle, which is included in Table \ref{tab1}.

\paragraph{Multi-physics simulation parameters}
The discharge was simulated for a full four AC-voltage cycles, to \SI{2e-3}{s}. To make sure that the quasi-steady state has been reached, only the last cycle was used for the time-average electric body force calculations. The flow field on the other hand was simulated up to \SI{5}{s}. 

\paragraph{Co-simulation parameters}

Co-simulation has been built on Comsol Multiphysics LiveLink for Simulink platform. An FMU block generated in Comsol is accessible via Simulink library browser. The FMU block built for this study accepts one input, \(V_{ap}\) and two outputs, the flowrate measured at the outlet of the amplifier system used for tracking and control, and the axial velocity at the center, which is used only for monitoring. There are three important parameters which are essential in safe and seamless communication of the two platforms without convergence issues. The communication step-size, which was set to \SI{1e-3}{s},  the steps of the time-dependent fluid solver which are stored for communication, and their relative interpolation tolerance which were set to \SI{1e-4}{s} and \(0.05\), respectively. The communication step-size is an option inherent to the Comsol FMU block for Simulink. Times to store, and their relative tolerance are accessible through Comsol model, but can be made available to set and change via the FMU block. The solver configuration for Simulink were set to variable step with initial step size of \SI{1e-10}{s}, maximum step size of \SI{1e-6}{s}, and relative tolerance of  \(1e-12\).

\begin{table}[htbp]
\caption{Parameters of the model under investigation.}
\begin{center}
\begin{tabular}{l l}

\hline
\hline     \\ [-5pt]                          
\textbf{Parameter} & \textbf{Value}  \\ [+2pt]  
\hline    \\ [-5pt]

Length of the converging nozzle, \(l_2\) & \SI{15}{mm}  \\
Length of the cylinders, \(l_3\) & \SI{40}{mm}  \\
Length of the diverging nozzle at zero angle, \(l_4\) & \SI{85}{mm}  \\
Radii of the inner cylinder, \(r_1\) & \(5, 10, 20\), and \SI{30}{mm}  \\
Air gap between the two cylinders, \(r_2\) & \SI{10}{mm}  \\
Jet-forming air gap downstream the two cylinders, \(l_1\) & \SI{7}{mm}  \\
Thickness of the body, \(l_{th}\) & \SI{5}{mm}  \\
Angles of the diverging nozzle, \(\alpha\) & \SI{0}{^{\circ}}, \SI{5}{^{\circ}}, and \SI{10}{^{\circ}}  \\
Width of the domain, \(l_w\) & \SI{200}{mm}  \\
Height of the domain, \(l_h\) & \SI{450}{mm}  \\ 
Length of the discharge electrodes,  \(l_d\)   & \SI{5}{mm} \\
Thickness of the discharge electrodes, \(D_d\) & \SI{50}{\mu m} \\
Length of the passive electrodes,  \(l_g\)   & \SI{5}{mm} \\
AC-voltage amplitude, \(V_{ap}\) &  \SI{14}{kV}  \\
AC-voltage frequency, \(f\) &  \SI{2}{kHz}  \\ 
Relative permittivity of the air, \(\epsilon_{r_a}\) & \(1.0006\)  \\
Relative permittivity of the dielectric material, \(\epsilon_r\) & \(4.0\)  \\[+5pt]

\hline     
\hline
\end{tabular}
\label{tab1}
\end{center}
\end{table}

\newpage
\subsection{Plasma and fluid flow simulation}

\subsubsection{Plasma simulation}

\begin{figure*}[htbp]
   \centering
    
     \begin{subfigure}[b]{0.48\textwidth}
    \includegraphics[width=1\textwidth]{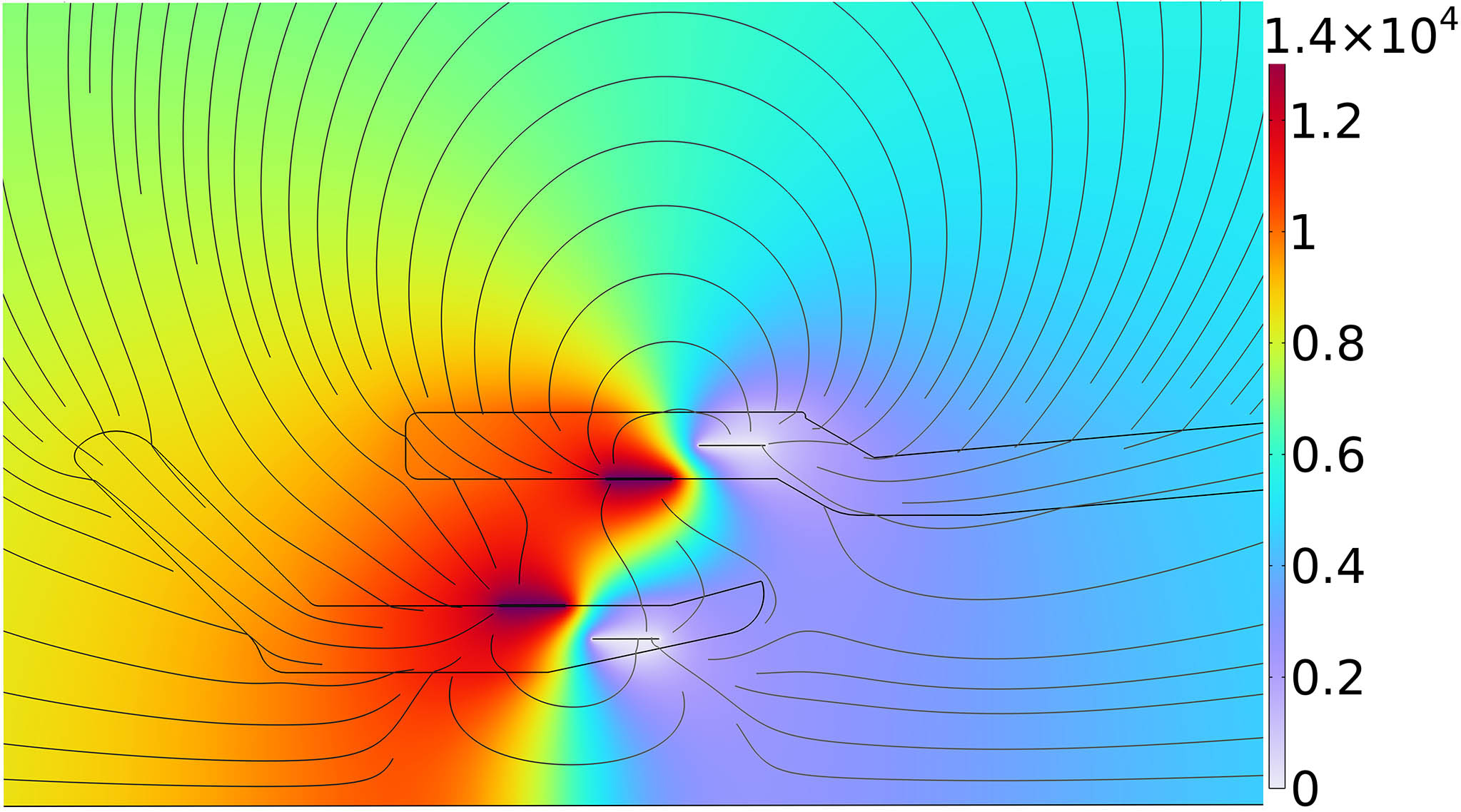}
    \caption{}
    \label{p1}
    \end{subfigure}
     \begin{subfigure}[b]{0.48\textwidth}
    \includegraphics[width=1\textwidth]{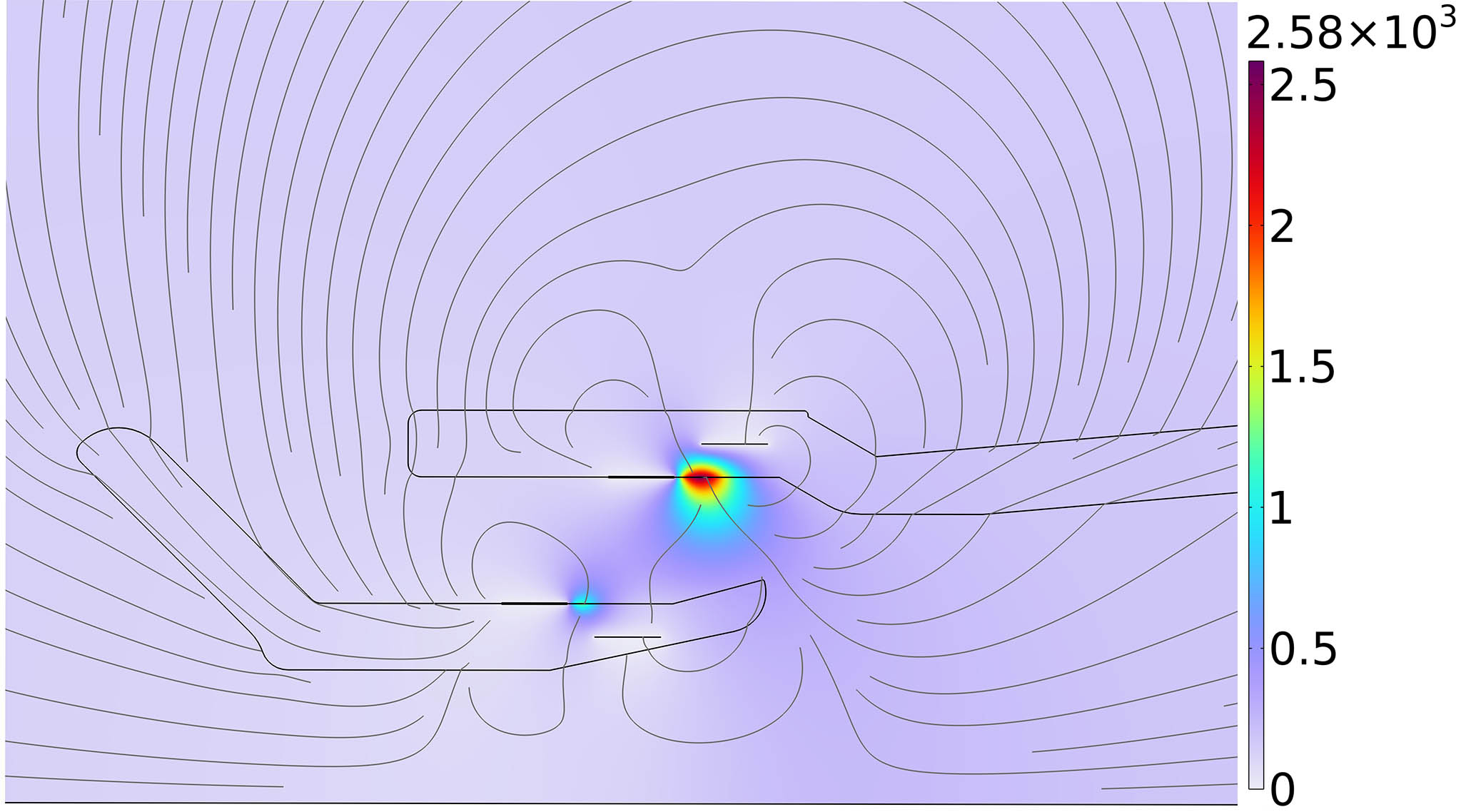}
    \caption{}
    \label{p2}
    \end{subfigure}
    
     \begin{subfigure}[b]{0.44\textwidth}
    \includegraphics[width=1\textwidth]{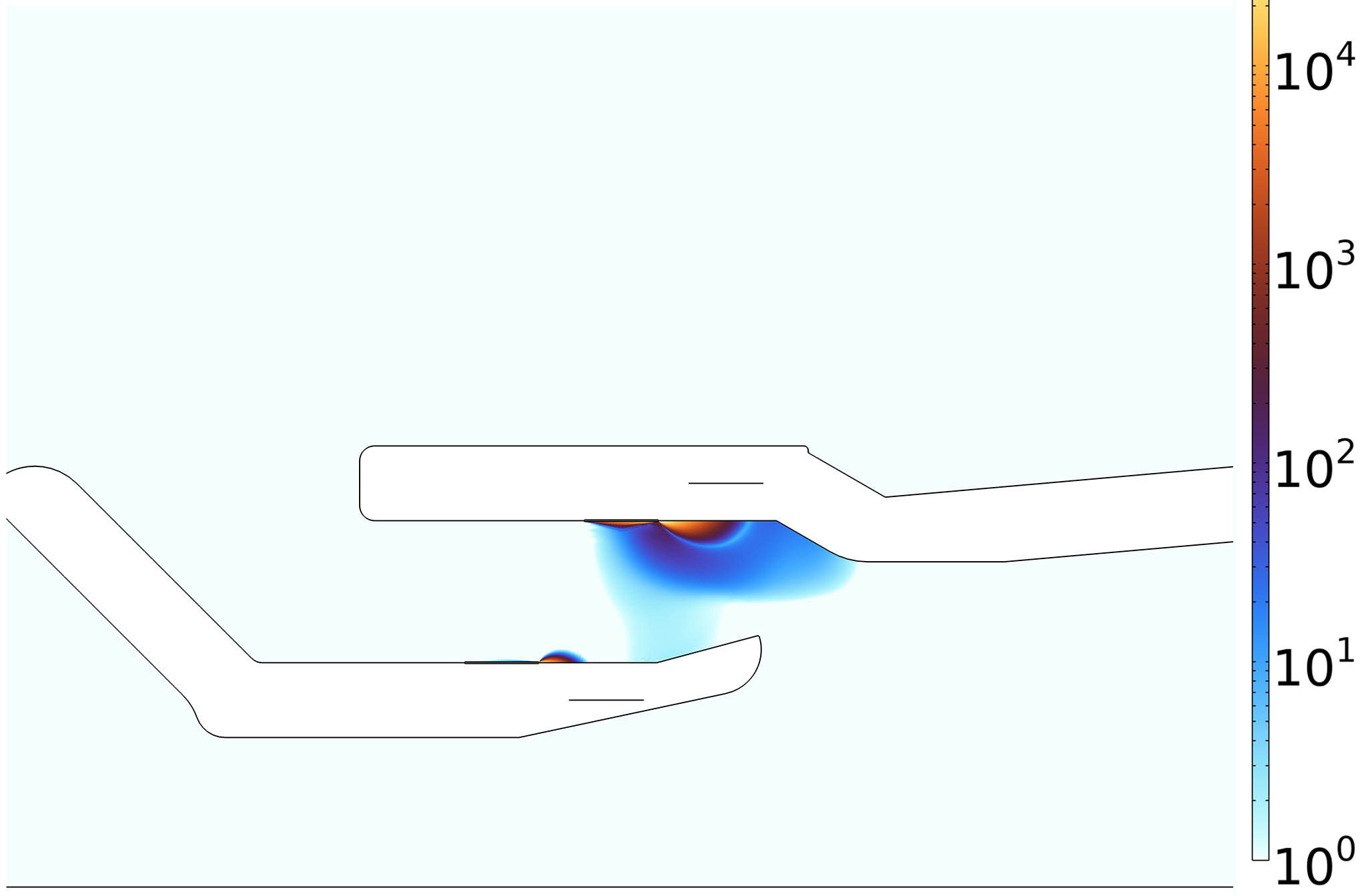}
    \caption{}
    \label{p3}
    \end{subfigure}
     \begin{subfigure}[b]{0.45\textwidth}
    \includegraphics[width=1\textwidth]{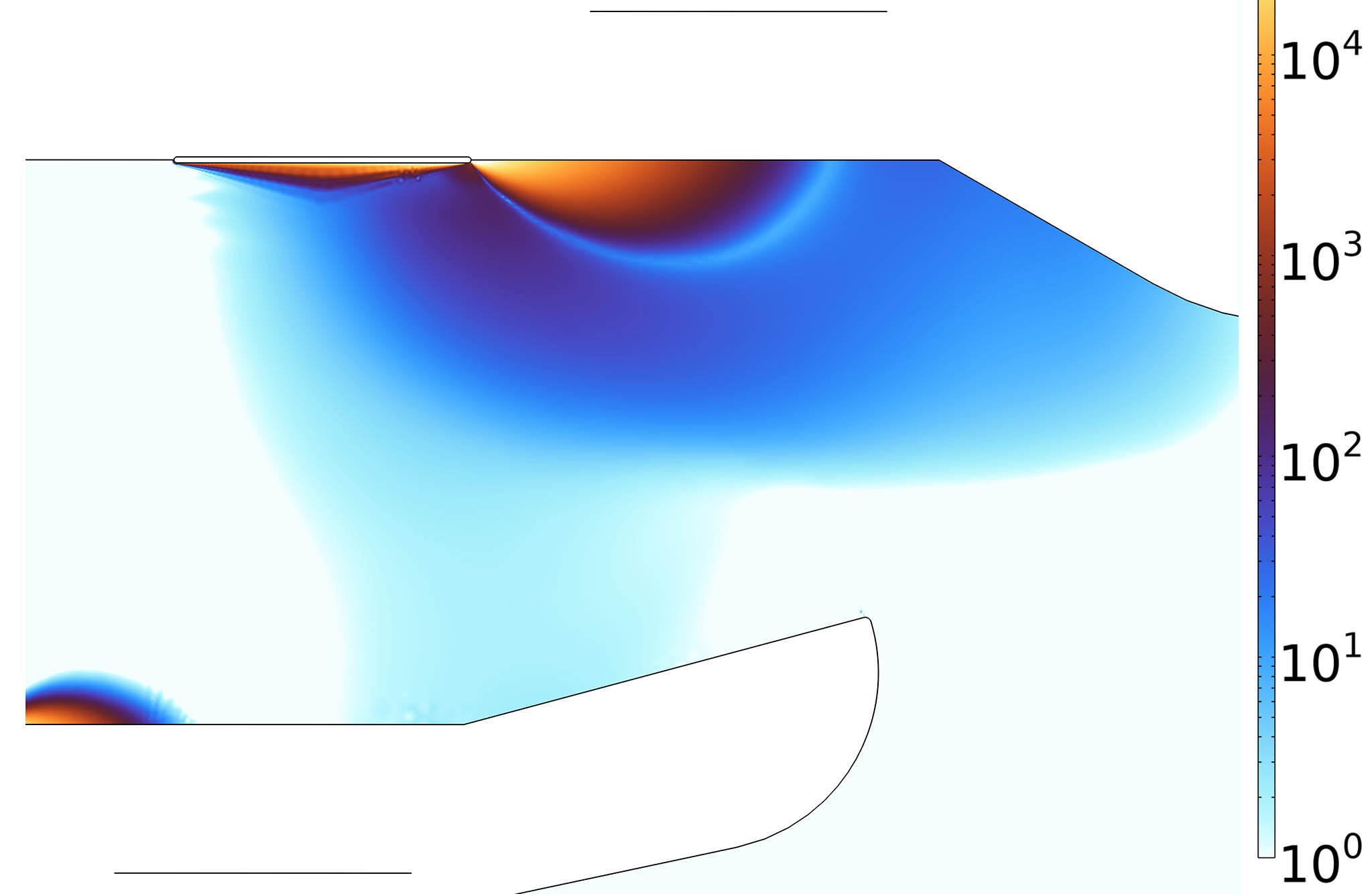}
    \caption{}
    \label{p4}
    \end{subfigure}
    
\caption{The electric field in \SI{}{Vm^{-1}} and the  electric field lines when the supply voltage is equal zero (\ref{p1}), or reaches its maximum value (\ref{p2}). The average electric body force \(F_{ehd}\) (\ref{p3}), and its magnified version in \SI{}{Nm^{-3}} (\ref{p4}).}
    \label{Plasma}
\end{figure*}

The fully coupled discharge equations were solved to obtain the space charge density and electric field formed in the space, when the sinusoidal voltage \(V_{ap}\) with frequency \(f\) is applied. The simulation starts at \(t=0\) and \(V_{ap}=0\), in the positive half-cycle of the AC voltage. The positive discharge initiates when the voltage increases and reaches a certain positive value. Positive charge species are injected into the space, with some charge depositing onto the dielectric surface. Increasing charge deposited on the surface hinders more injection of charge from the discharge electrode, and the discharge stops before the voltage goes to the negative half-cycle, and this process cycles on. When the voltage hits the maximum value (i.e. \(t= nT+ \frac{pi}{2f} \), where \(n\) is an integer and \(T\) is the period of the input voltage), the electric field and potential formed by the discharge process are depicted in figure \ref{p1}. When the voltage hits zero after the positive cycle, both discharge electrodes are at zero potentials. However, the positive charge on the surface of the dielectric, as well as the space charge remainders in the space, produce electric potential and field, which are shown in figure \ref{p2}. The space charge density and electric field give the EHD body force \(F_{ehd}\), shown in figures \ref{p3} and \ref{p4}. The force is concentrated and most intensified near the tip of the discharge electrodes. This is not surprising to see that the force generated by the discharge mounted on the outer cylinder is stronger. It is axially located downstream the other discharge electrode, and affects the discharge behavior. Utilizing two sets of actuators facilitates the generation of more directed plasma-induced jets and minimizes flow separation. For optimal results, the supplied voltages to the actuators can be individually tailored to further reduce flow separation and the generation of undesirable yet inevitable vortices. However, this aspect falls outside the scope of this study, with the supplied voltages assumed to be equal for both actuators. Placing the actuators too closely either radially or axially can lead to diminishing discharge, thereby reducing performance. Conversely, if they are spaced too far radially due to a significant size difference between the outer and inner cylinders, the induced jet may not form adequately to improve the Coanda jet. A more thorough examination of the impact of electric field distribution, varying dielectric permittivity, and surface conductivity values on plasma jets also remains beyond the scope of this study.

\newpage
\subsubsection{Fluid flow simulation}

\begin{figure*}[htbp] 
   \centering
    
     \begin{subfigure}[b]{0.48\textwidth}
    \includegraphics[width=1\textwidth]{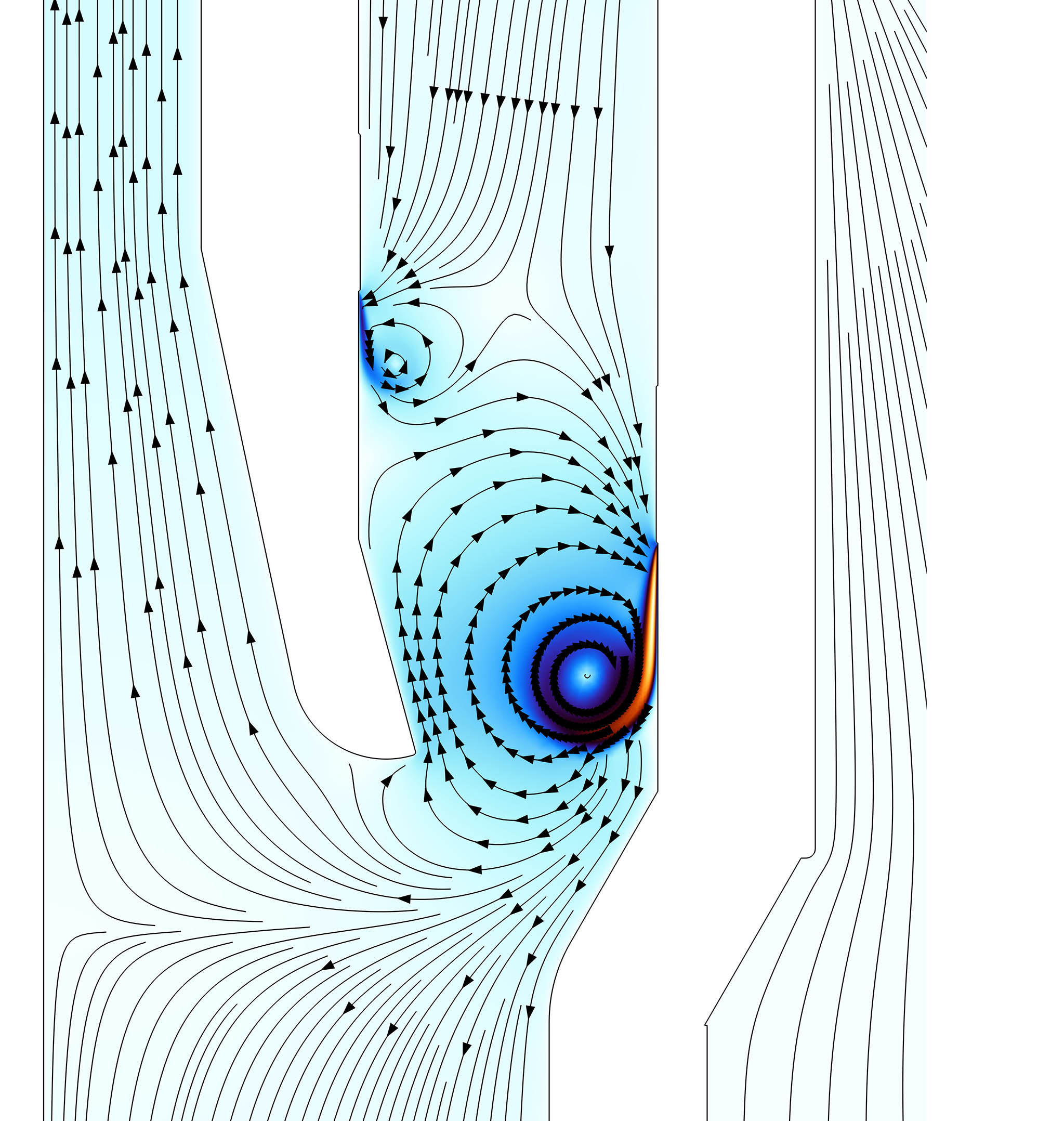}
    \caption{}
    \label{v1}
    \end{subfigure}
    %\hfill
     \begin{subfigure}[b]{0.48\textwidth}
    \includegraphics[width=1\textwidth]{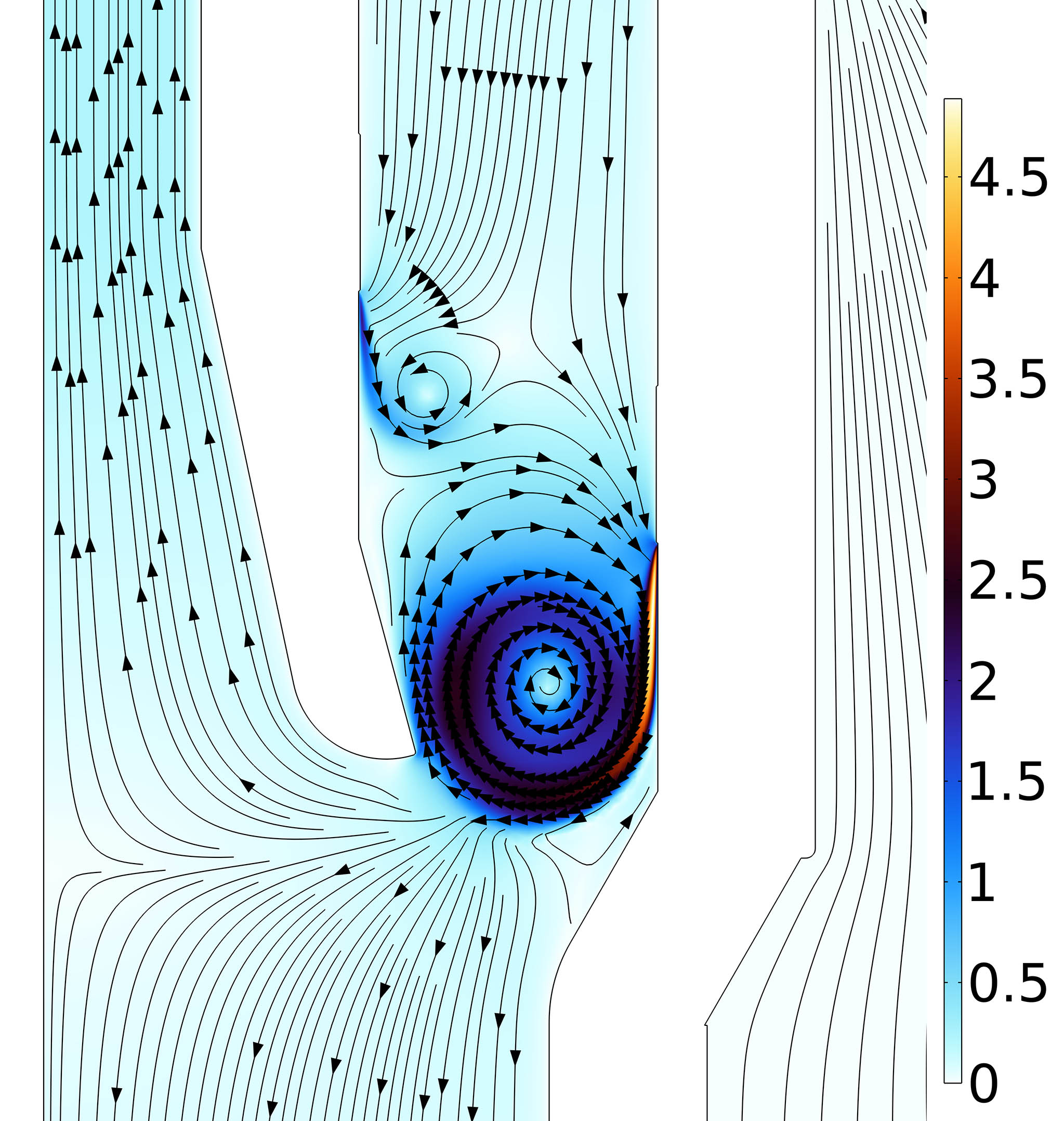}
    \caption{}
    \label{v2}
    \end{subfigure}
   % \hfill

\caption{Vortex formation following the activation of the actuator at times:  \(t=0.005\) (\ref{v1}), and  \(t=0.015\) (\ref{v2}). Velocities are in \SI{}{ms^{-1}}.}
    \label{V}
\end{figure*}

\begin{figure*}[htbp]
   \centering
    
     \begin{subfigure}[b]{0.48\textwidth}
    \includegraphics[width=1\textwidth]{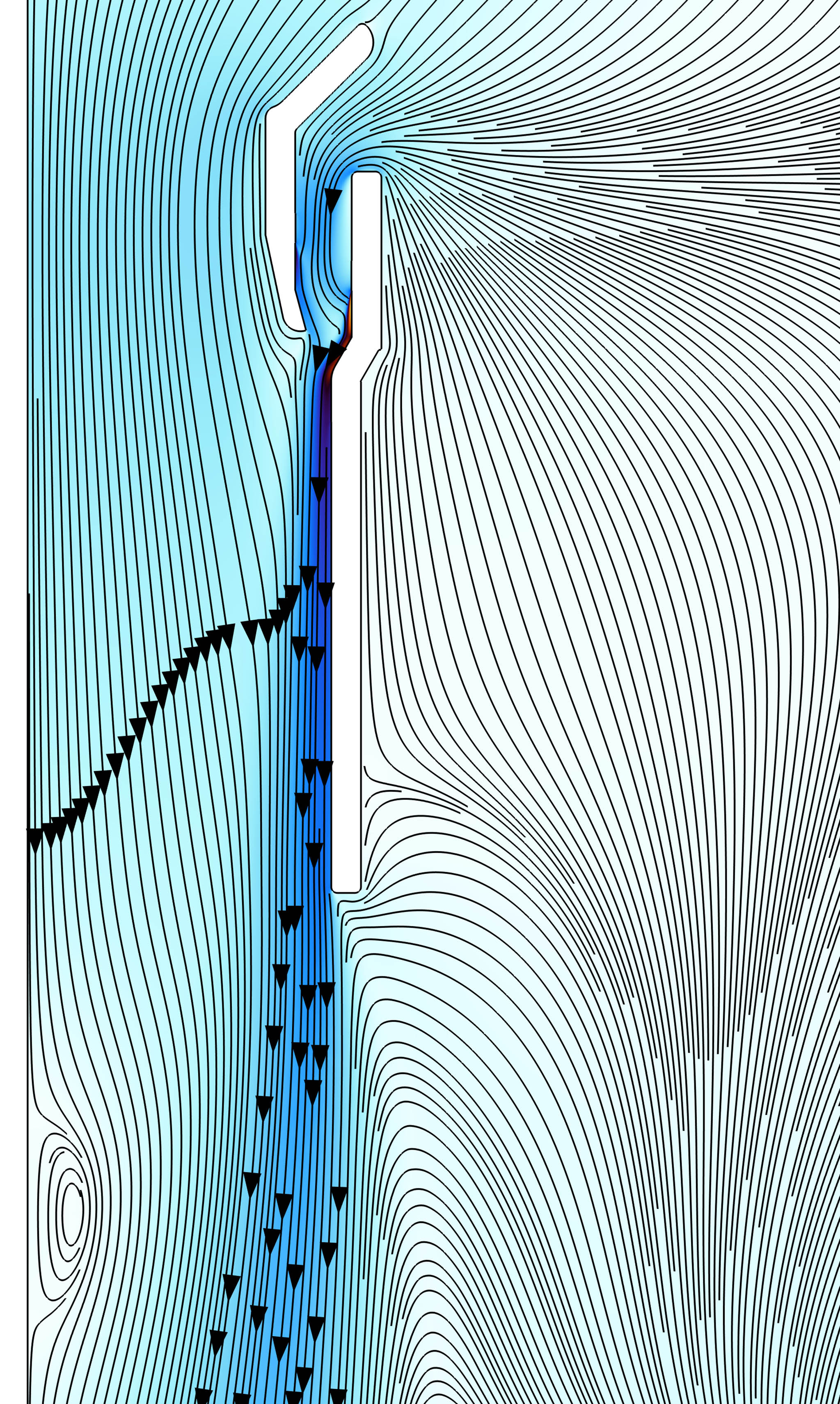}
    \caption{}
    \label{w1}
    \end{subfigure}
    %\hfill
     \begin{subfigure}[b]{0.48\textwidth}
    \includegraphics[width=1\textwidth]{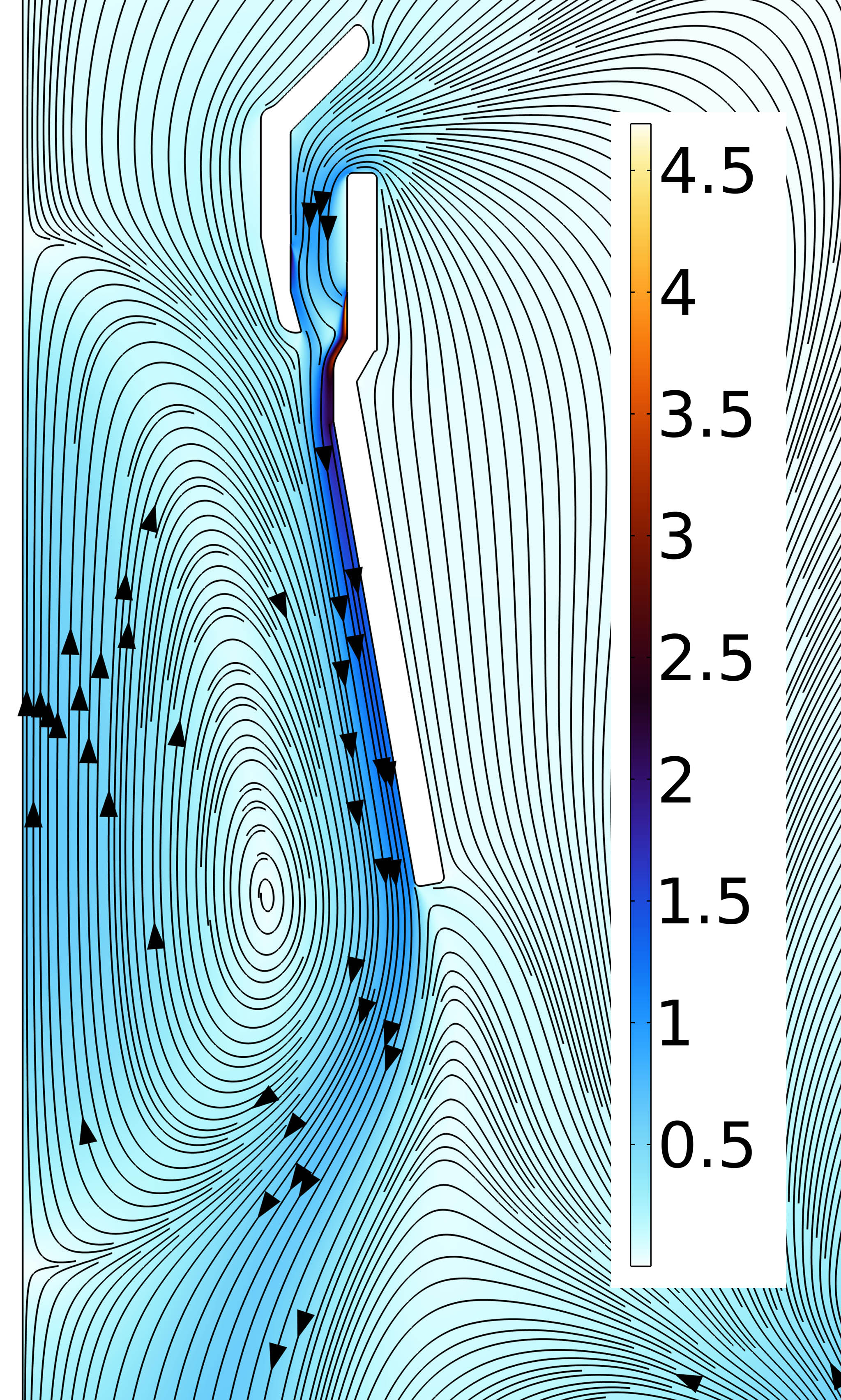}
    \caption{}
    \label{w2}
    \end{subfigure}
   % \hfill

\caption{Air flow field and streamlines shown for \(r_1= \SI{40}{mm}\), and \(\alpha=\SI{0}{^{\circ}}\) (\ref{w1}), and  \(\alpha=\SI{10}{^{\circ}}\) (\ref{w2}). Velocities are in \SI{}{ms^{-1}}.}
    \label{W}
\end{figure*}

\begin{figure*}[htbp]
   \centering
    
     \begin{subfigure}[b]{0.32\textwidth}
    \includegraphics[width=1\textwidth]{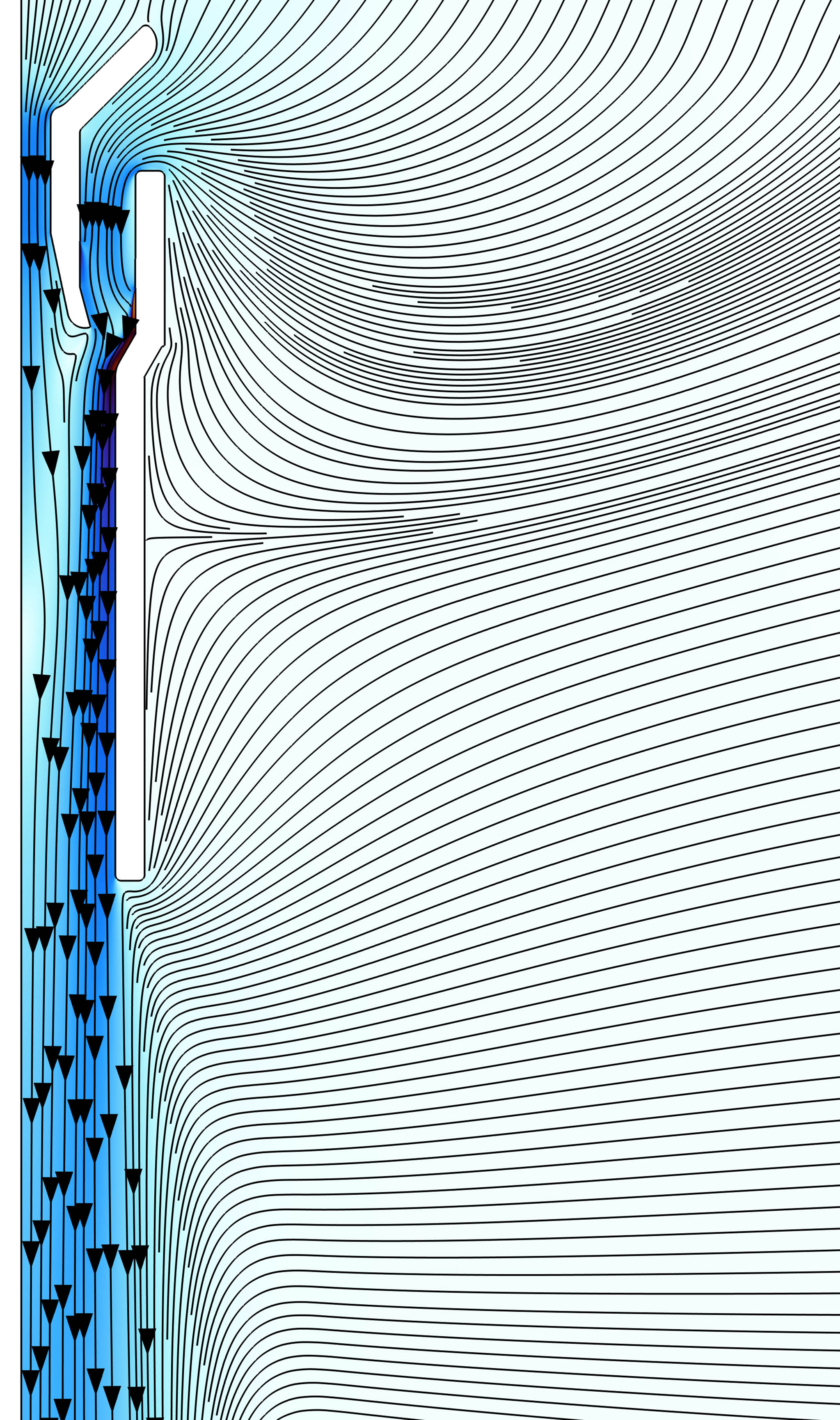}
    \caption{}
    \label{s1}
    \end{subfigure}
     \begin{subfigure}[b]{0.32\textwidth}
    \includegraphics[width=1\textwidth]{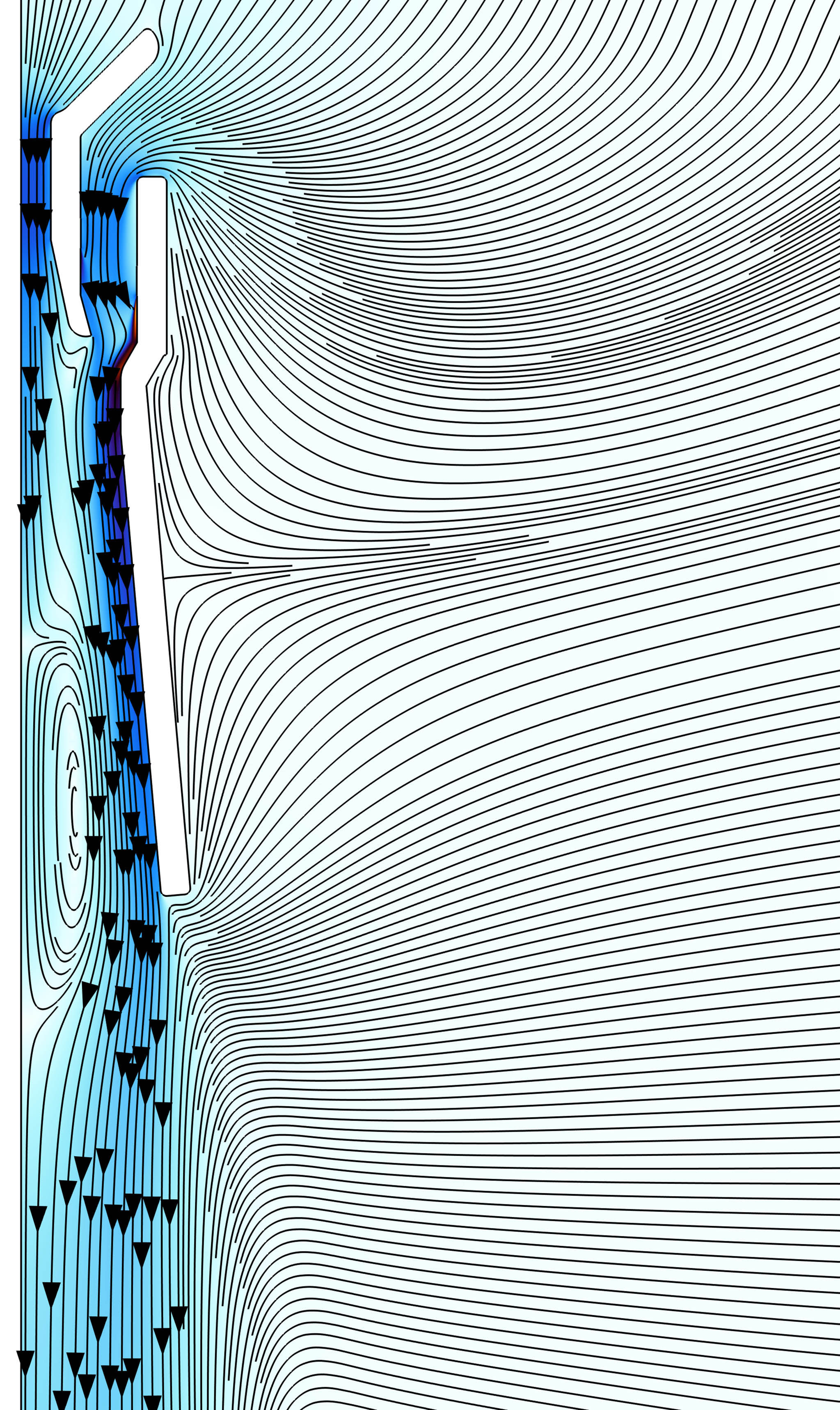}
    \caption{}
    \label{s2}
    \end{subfigure}
     \begin{subfigure}[b]{0.32\textwidth}
    \includegraphics[width=1\textwidth]{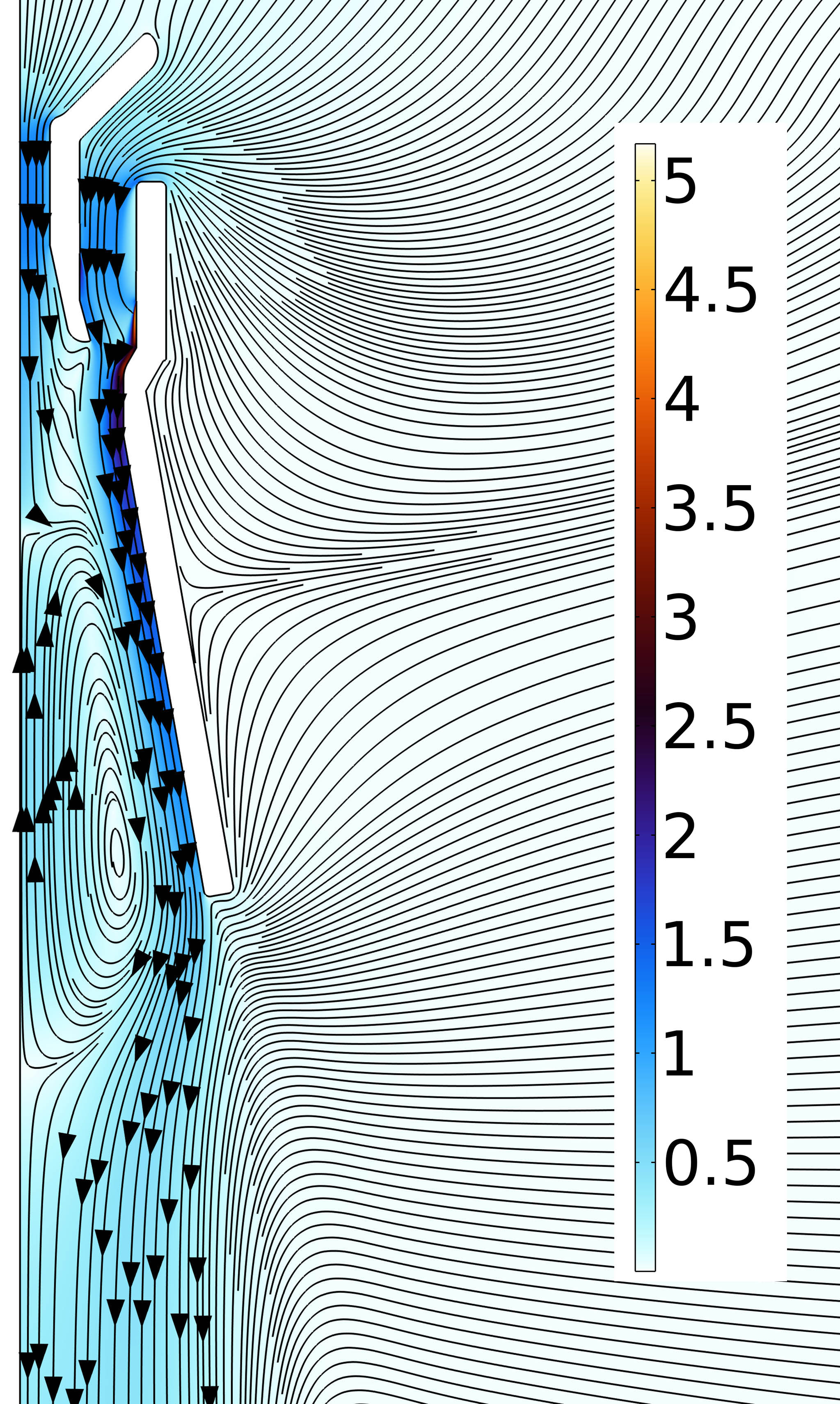}
    \caption{}
    \label{s3}
    \end{subfigure}

     \begin{subfigure}[b]{0.32\textwidth}
    \includegraphics[width=1\textwidth]{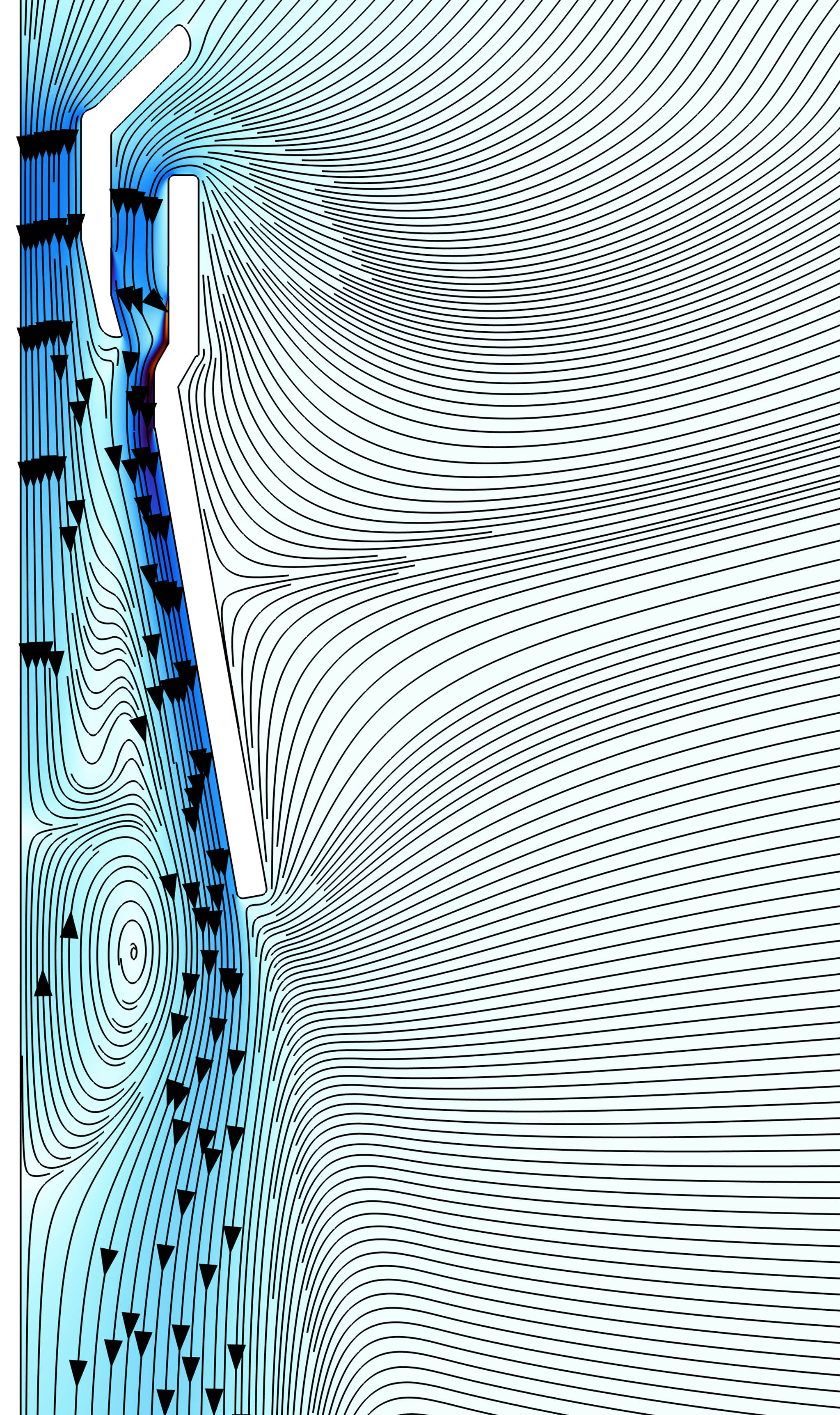}
    \caption{}
    \label{s4}
    \end{subfigure}
     \begin{subfigure}[b]{0.32\textwidth}
    \includegraphics[width=1\textwidth]{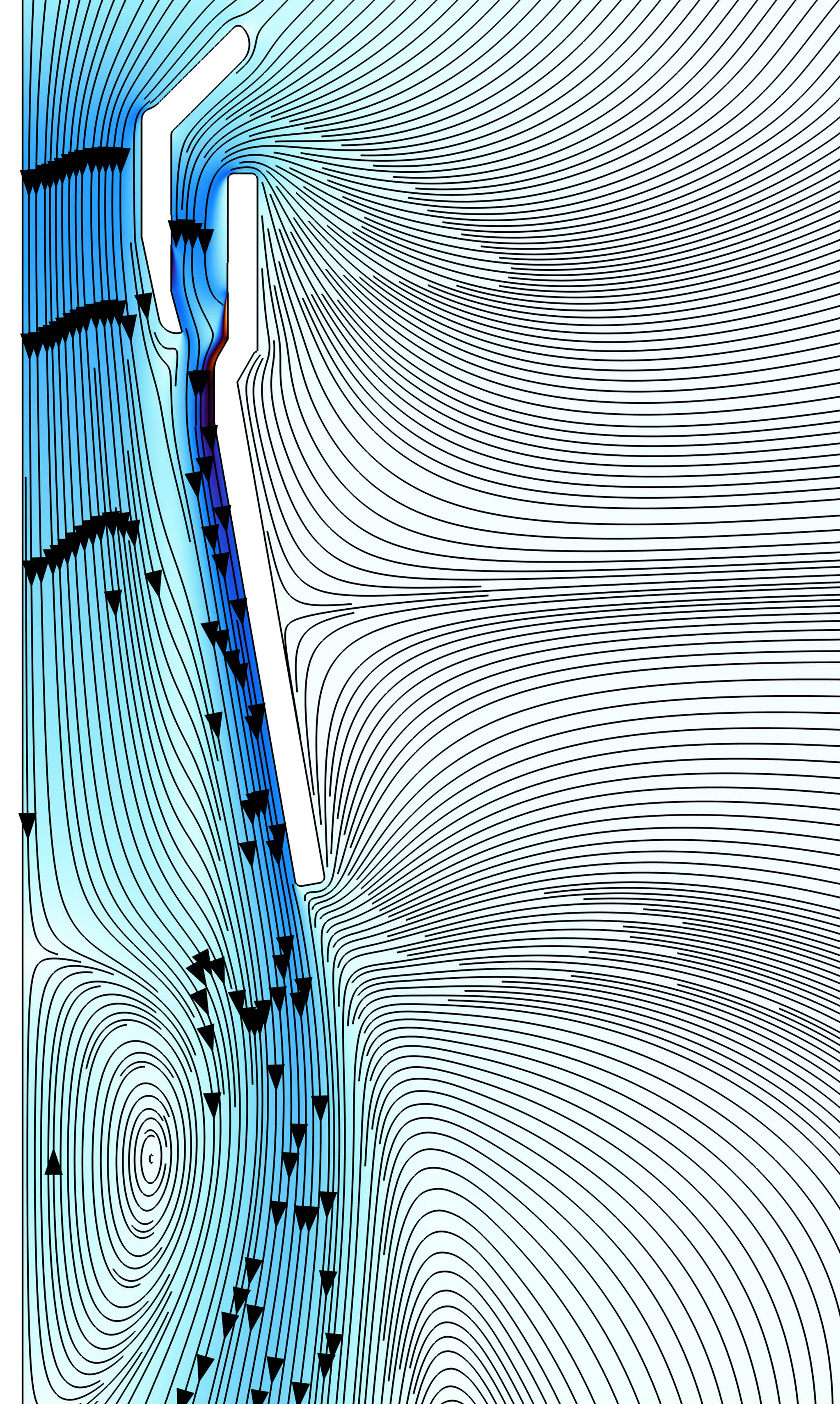}
    \caption{}
    \label{s5}
    \end{subfigure}
     \begin{subfigure}[b]{0.32\textwidth}
    \includegraphics[width=1\textwidth]{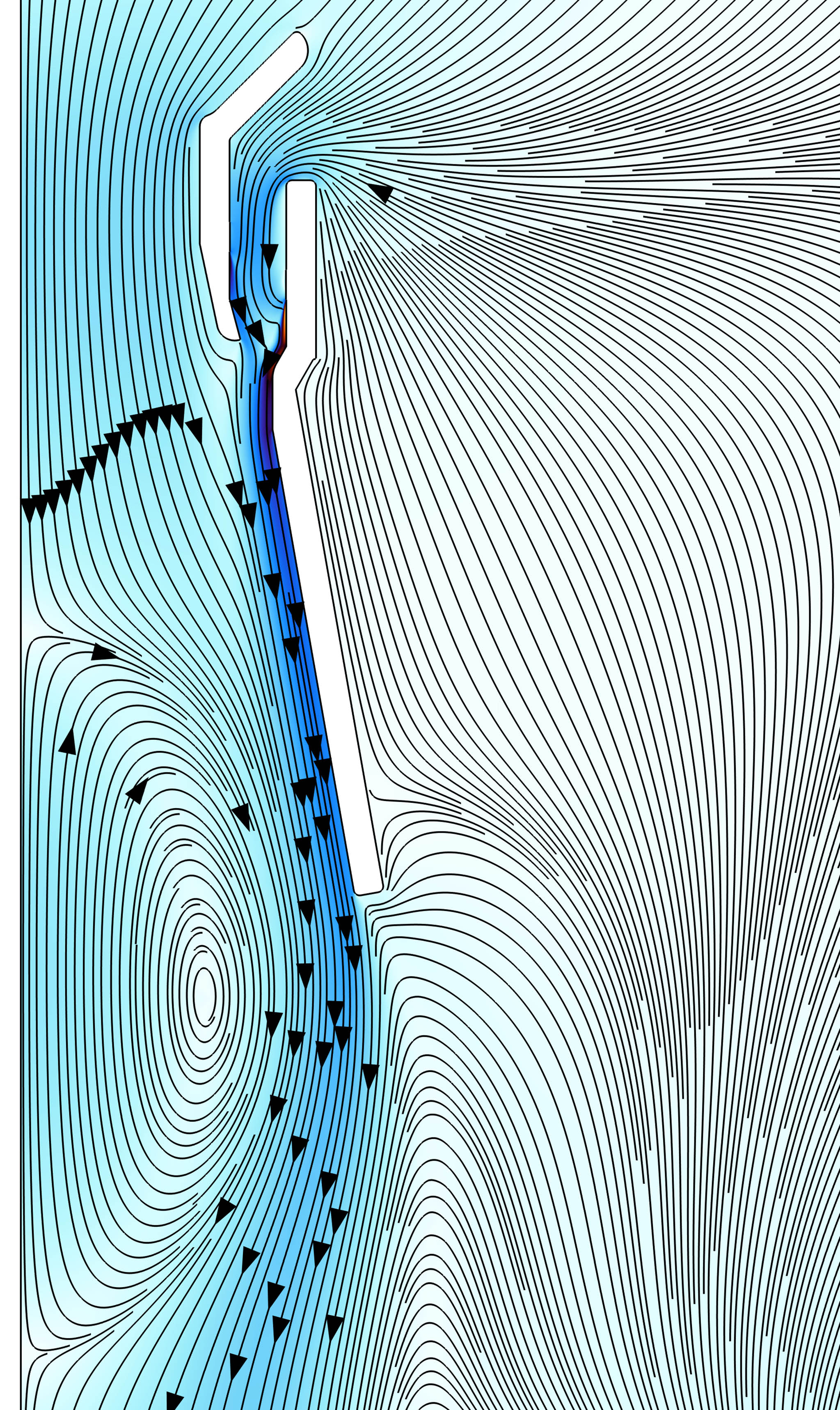}
    \caption{}
    \label{s6}
    \end{subfigure}
    
\caption{Air flow field and streamlines presented for \(r_1=\SI{5}{mm}\), and \( \alpha=\SI{0}{^{\circ}}, \SI{5}{^{\circ}}\), and \(\SI{10}{^{\circ}}\) in \ref{s1}, \ref{s2}, and  \ref{s3}, respectively. For \(\alpha=\SI{10}{^{\circ}}\) and \(r_1=\SI{10}{mm}\), \(\SI{20}{mm}\) and \(\SI{30}{mm}\), results are presented in \ref{s4}, \ref{s5}, and  \ref{s6}, respectively.}
    \label{S}
\end{figure*}

The time-averaged electric body force obtained from the plasma simulation has been incorporated into the NS equations for the fluid flow simulation using the \(k-\omega\) turbulence model. The simulation starts at \(t=0\) when the body force activates and affects the flow field, by forming vortices in the discharge regions. Vortices are identifiable by the streamlines overlaid on the velocity magnitude maps in figures \ref{V} \ref{W} and \ref{S}. Initially, two vortices are formed, and captured at \(t=0.005\), and \(t=0.015\) in figures \ref{v1} and \ref{v2}, respectively. This is known that the electric body force generated by DBD actuators form a low-pressure area above the discharge electrode, which pulls the flow towards the surface of the dielectric \cite{shaygani2021dielectric}, before the body force pushes forward in the axial direction to eventually form a vortex. This phenomenon is clearly captured by the streamlines shown in the figures. The vortex generated by the stronger body force is larger and stronger, and eventually forms the air jet, which sticks  to the wall for some time due to the Coanda effect.

The flow field simulated for a device with \(r_1=\SI{30}{mm}\) and \(\alpha=\SI{0}{^{\circ}}\), and \(\SI{15}{^{\circ}}\) are shown in figures \ref{w1} and \ref{w2}, respectively. 

Depending on the geometry and different parameters such as \(r_1, r_2, \alpha\), etc., the Coanda jet formed by the actuator which sticks to the wall until it leaves the device, can form a stream of flow without (\ref{w1}), or with vortices developed inside the diverging nozzle (\ref{w2}). Such vortices influence the dynamics of the flow, and therefore the performance of the actuation as an amplification method. This is known that such an EHD flow control configuration amplifies the flow formed by the actuator, by enforcing the other stream of flow coming from the central inlet to merge in, and exit from the outlet \cite{rubinetti2023electrohydrodynamic,rubinetti2023silico}. Our findings suggest that this can get quite complicated with formation of the vortices. To achieve the desired behavior and performance, it is crucial to engage in meticulous design and carefully consider the parameters involved.

Different behaviors emerging from choosing different parameters are presented in figure \ref{S}. Setting \(r_1=\SI{5}{mm}\), and \( \alpha=\SI{0}{^{\circ}}, \SI{5}{^{\circ}}\), and \(\SI{10}{^{\circ}}\), the streamlines are shown in figures \ref{s1}, \ref{s2}, and \ref{s3}, respectively. Keeping \( \alpha=\SI{10}{^{\circ}}\), the streamlines are shown in figures \ref{s4}, \ref{s5}, and \ref{s6} for \(r_1=\SI{10}{mm}, \SI{20}{mm}\) and \(r_1=\SI{30}{mm}\), respectively. When \(\alpha\) is nonzero (all except \ref{s1}), the vortices generated in the diverging nozzle force the air flow to move in the opposite direction. Increasing \(r_1\) and \( \alpha\) causes the formation of larger vortices. When \(\alpha=\SI{0}{^{\circ}}\), no vortex is formed, and the flow amplification occurs without a reverse flow. Even if there is a reverse flow, amplification happens since the backward flow rate is lower than the flow rate entering through \(inlet_1\).

This is worth emphasizing that there are other parameters too, which influence how the system behaves,including the applied voltage and frequency supplied to the actuator, and other geometrical parameters in addition to \(\alpha\). Amplification with different efficiencies may be achieved by carefully selecting other arrangements of parameters, which is problem-specific, and out of the scope of this study. All simulation results presented in figures \ref{W}, and \ref{S}, were captured at \(t=\SI{5}{s}\), when the flow has reached to a steady state.

\newpage
\subsection{System identification and control design}

The actuated flow system with a modulated input voltage, and measured outflow rate as output was considered to design a control system. Six cases have been compared: \(\alpha =\SI{10}{^{\circ}}\) with \( r_1=\SI{10}{mm}, \SI{20}{mm}\), and \(\SI{30}{mm}\) as Case 1 to 3, respectively, and \(r_1=\SI{5}{mm}\) with \(\alpha =    \SI{0}{^{\circ}}, \SI{5}{^{\circ}} \), and \(\SI{10}{^{\circ}}\) as Case 4 to 6, respectively.

\begin{figure*}[htbp]
   \centering
    
     \begin{subfigure}[b]{0.90\textwidth}
    \includegraphics[width=1\textwidth]{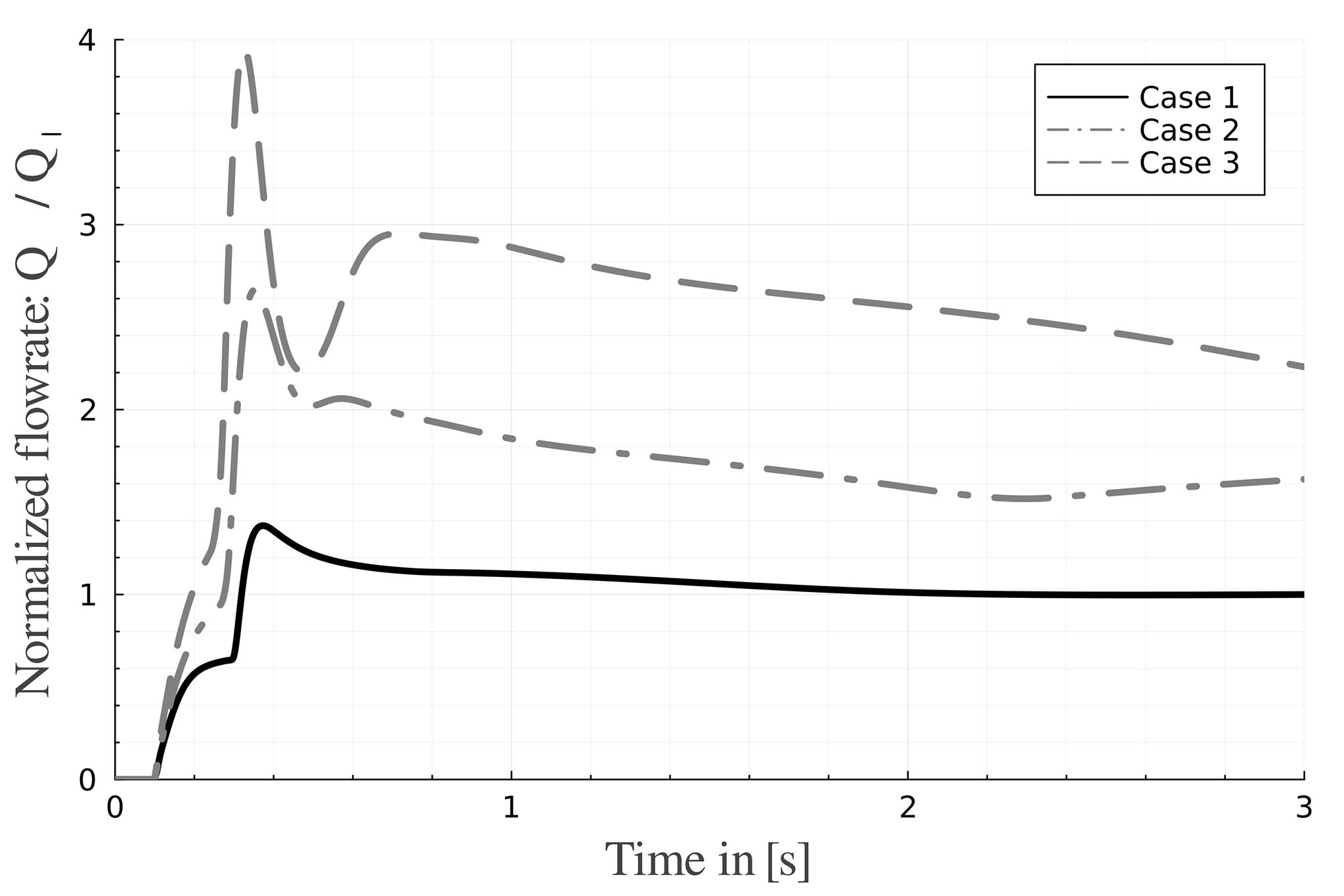} 
    \caption{}
    \label{tf1}
    \end{subfigure}
     \begin{subfigure}[b]{0.90\textwidth}
    \includegraphics[width=1\textwidth]{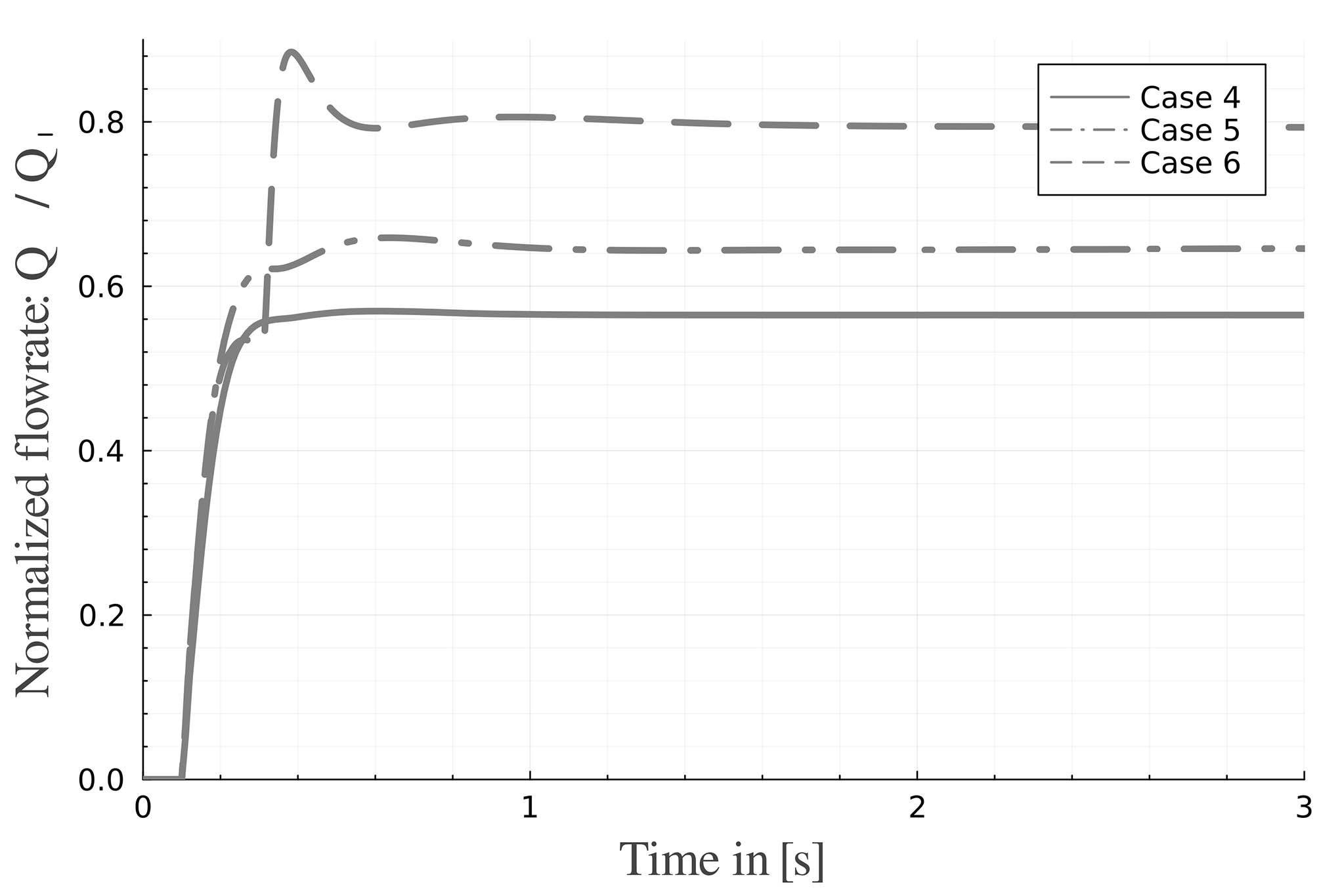} 
    \caption{}
    \label{tf2}
    \end{subfigure}
    
    \caption{The normalized flow rates at the outlet, when an input step is applied to the actuator.}
    \vspace{-0.1in}
    \label{tf}
\end{figure*}

Assuming \(V_{ap}\) a unit input, a unit input modulation at time \(t=\SI{0.1}{s}\), the outflow rates normalized by the steady flow rate of Case 1 and have been compared in figure \ref{tf}. Among the cases we examined, as \(\alpha\) increases from \SI{0}{^{\circ}} to \SI{10}{^{\circ}}, the steady state outflow rates increase. The same trend was observed when increasing \(r_1\). Increasing \(r_1\) or \(\alpha\) increases the response time of the system. The response time can be attributed to the vortices developed inside the device, before it reaches the steady state.

\vspace{+0.1in}

Considering Case 1 as our main case, the full order model (FOM) system can be approximated with the following seventh order system of the form:

\begin{equation}\label{tfeq}
\scriptsize
F(s) =  G \frac{1.317 s^6 + 1194 s^5 -9834 s^4 + 3.348 \times 10^6 s^3 - 1.225 \times 10^7 s^2 + 9.581 \times 10^8 s + 5.962 \times 10^8}{s^7 + 53.1 s^6 + 5089 s^5 + 1.621 \times 10^5 s^4 + 5.59 \times 10^6 s^3 + 8.102 \times 10^7 s^2 + 8.433 \times 10^8 s + 6.245 \times 10^8}
\end{equation}

where \(G\) scales the unit response to the output flow rate. The parameters of the control strategies were calibrated on (\ref{tfeq}) and tested on the FOM. A Two-degree-of-freedom (2DOF) PID controller with the following control law has been used as a baseline for performance test of the ADRC controller:

\begin{equation}\label{PID}
u =   P (b r - y)  +  I (r-y)  +  D  \frac{N}{1+N/s}  (c r - y)
\end{equation}
where \(P,\; I,\; D,\; N,\; b\), and \(c\) are model parameters, and \(r,\; y\), and \(u\) are reference signal, system output, and system input, respectively. Calibrated parameters are summarized in Table \ref{tab2}.  

\begin{figure*}[htbp]
   \centering
    
     \begin{subfigure}[b]{0.90\textwidth}
    \includegraphics[width=1\textwidth]{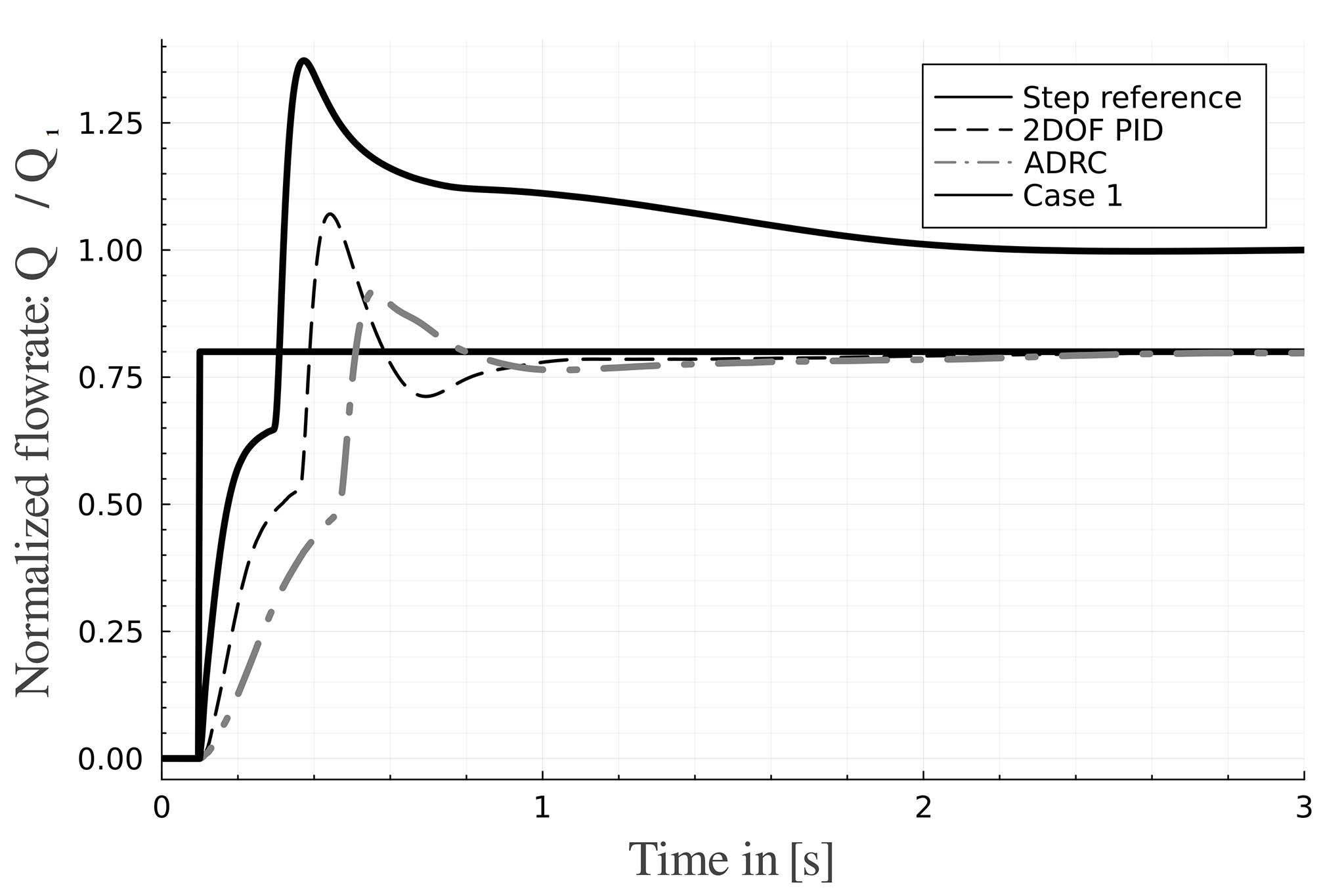} 
    \end{subfigure}
    
    \caption{Closed-loop step responses, ADRC vs 2DOF PID.   Responses are normalized flow rates at the outlet, when an input step is applied to the actuator.}
    \vspace{-0.1in}
    \label{step}
\end{figure*}

\begin{figure*}[htbp]
   \centering
    
     \begin{subfigure}[b]{0.90\textwidth}
    \includegraphics[width=1\textwidth]{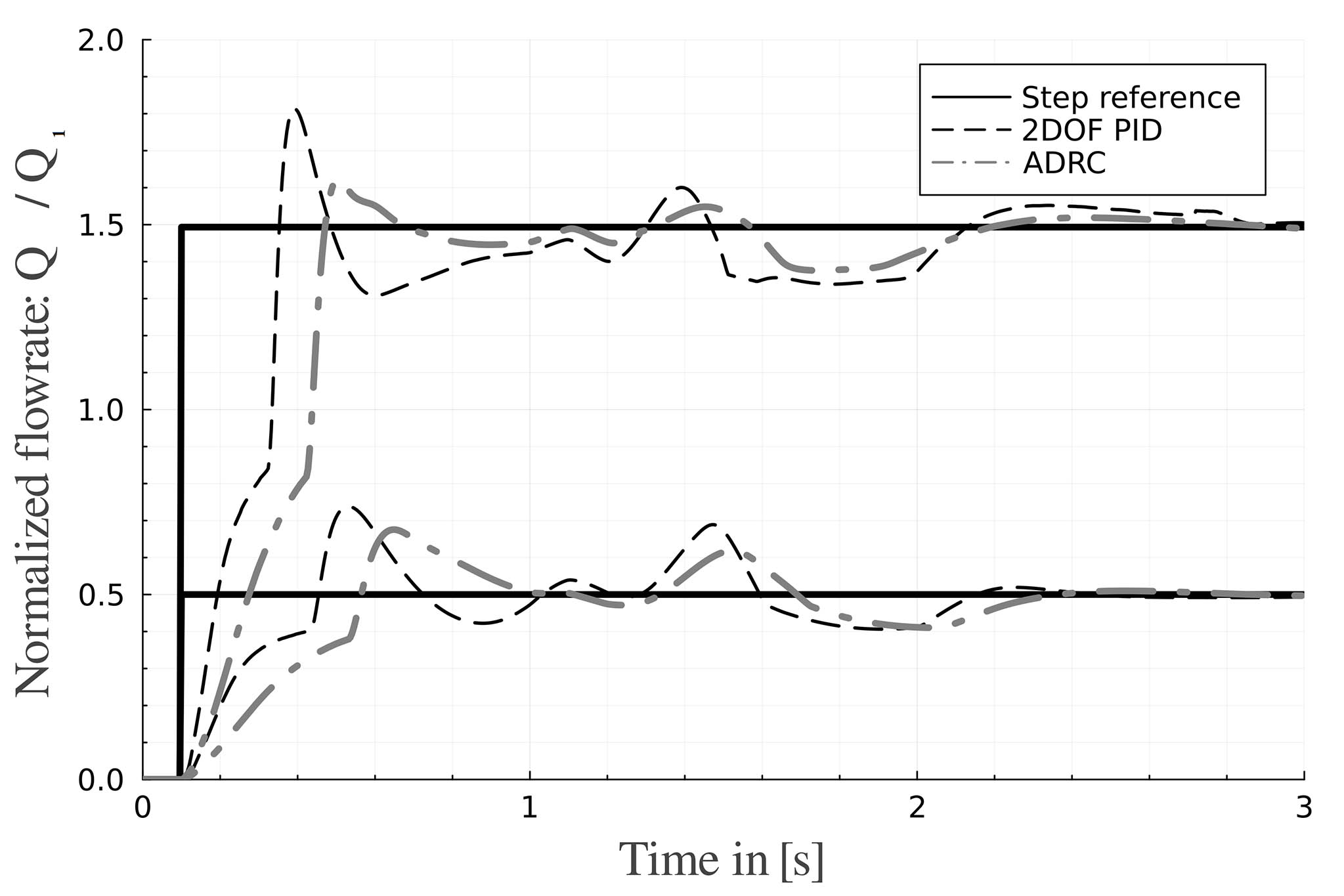} 
    \end{subfigure}
    
    \caption{Disturbance rejection comparison, ADRC vs 2DOF PID strategies. Responses are normalized flow rates at the outlet. A disturbance is introduced at \(t=\SI{0.5}{s}\), while an input step has been applied to the actuator.}
    \vspace{-0.1in}
    \label{disturbance}
\end{figure*}

The performance of the ADRC control law was assessed by evaluating its closed-loop step response and disturbance rejection capabilities, which were then compared to the baseline performance, presented in figures \ref{step}, and \ref{disturbance}, respectively.

The control parameters are tuned to have less than \(25\%\) overshoot, reach the fastest settling and rise times, while maintaining the stability of the system. In addition, convergence of the co-simulation must be maintained, which serves as additional system constraints. Certain parameter configurations may cause the co-simulation system to diverge due to rapid changes in the control command. It is essential to carefully set these parameters to ensure stability and prevent unpredictable behaviors. Such parameter configurations were automatically disregarded during numerical experimentation.

A disturbance was introduced to the air flow simulation to test the performance of the control systems in the presence of external disturbances and uncertainties unaccounted for in the identified system. As mentioned before, zero normal stress is assumed on the boundaries of the flow simulation. The disturbance was introduced as normal pressure fluctuations on the boundaries upstream of the device at time \(t = \SI{1.0}{s}\), in a form of a bipolar pulse with the duration of \(t = \SI{0.2}{s}\), and amplitude of \(\SI{0.15}{N m^{-1}}\). Figure \ref{disturbance} illustrates the superior performance of ADRC compared to the PID controller, in the presence of pressure disturbances, for two reference inputs.

\vspace{+0.1in}

\vspace{+0.1in}

\begin{table}[htbp]
\caption{Parameters of the investigated control strategies.}
\begin{center}
\begin{tabular}{l l l l}

\hline
\hline     \\ [-5pt]

\multicolumn{2}{c}{\textbf{ADRC}} & \multicolumn{2}{c}{\textbf{2DOF PID}}   \\

Parameter & Value & Parameter & Value \\ [+2pt]  
\hline    \\ [-5pt]

\(b_0\)        & \(3e+1\)   &  \(P\)   & \(0.512/G\)   \\  
\(\beta_{o1}\) & \(7.5e+2\)   &  \(I\)   & \(5.58/G\)  \\
\(\beta_{o2}\) & \(1.875e+5\)   &  \(D\)   & \(-0.018/G\)   \\
\(\beta_{o3}\) & \(1.562e+1\)  &  \(N\)   & \(28.868\)  \\
\(\alpha_1\)   & \(0.85\)  &  \(b\)   & \(0.893\)   \\
\(\alpha_2\)   & \(0.15\)  &  \(c\)   & \(0.893\)   \\
\(\delta\)     & \(8e-2\)  &          &              \\
\(R\)          & \(1e+5\)  &          &              \\
\(\beta_1\)    & \(1e+2\)   &          &              \\
\(\beta_2\)    & \(1\)   &          &              \\
  
  \\[+5pt]

\hline     
\hline
\end{tabular}
\label{tab2}
\end{center}
\end{table}

\newpage
\section{Conclusions}

In this investigation, we designed and analyzed a Dielectric Barrier Discharge (DBD) plasma actuator for controlling an airflow control device. The actuator utilized electrohydrodynamic (EHD) body force to induce an air jet within the air gap between two concentric cylinders, leading to the creation of a suction area through the Coanda effect upon passing through the diverging nozzle. The control of the outflow rate was achieved by the plasma actuator using the Active Disturbance Rejection Control (ADRC) strategy, which was compared to the conventional PID controller.

The air amplifier system with the plasma actuator was modeled using a co-simulation approach, seamlessly linking COMSOL multi-physics for the CFD simulation of plasma and airflow, and Simulink for the implementation of the control law. The two-species model of discharge was employed for the DBD plasma simulation, and the electric body force from the plasma simulation was utilized in the Navier-Stokes equations for the turbulent EHD flow simulation. The control design was based on input (actuator voltage) and output (outlet flow rate) data, enabling the analysis of the plasma-air flow system.

The numerical co-simulation demonstrated the effectiveness of ADRC in enhancing air control dynamics compared to the PID controller. ADRC exhibited superior performance in reference tracking and noise rejection, effectively handling flow fluctuations and sudden pressure changes induced synthetically on the boundaries upstream of the air actuator device.

Building upon this, the study broadens its understanding by delving into the impact of various parameters on flow rate. It also investigates closed-loop control for achieving consistent airflow regulation. While the primary focus has been on numerical methods, this study sets the stage for potential experimental validations in the future. It provides a foundational framework that subsequent research can leverage, offering valuable insights for both the research community and practicing engineers in the domain of airflow dynamics control.

In conclusion, the present study successfully developed and analyzed a DBD plasma actuator-based airflow control system, integrating the capabilities of EHD forces and Coanda effect. The co-simulation approach allowed for a comprehensive understanding of the system's behavior and performance with and without control strategies. The findings highlight the superiority of ADRC over PID control in achieving improved airflow control and noise rejection, contributing to the advancement of plasma actuation techniques for flow control applications.

\hfill
\newpage

\bibliography{myref}
\end{document}